\pdfoutput=1

\documentclass[twocolumn,prb]{revtex4-1}
\usepackage{amssymb,amsmath,bm}
\usepackage{graphicx}%
\usepackage{dcolumn}
\usepackage{color}
\usepackage{longtable}
\usepackage[dvips]{epsfig}
\usepackage[perpage,symbol*]{footmisc}
\usepackage{multirow}
\usepackage{verbatim}
\usepackage{tablefootnote,threeparttable}
\usepackage{braket}

\usepackage{simpler-wick} 

\usepackage{filecontents} 

\usepackage{xr}

\newcommand*{\citen}{}
\DeclareRobustCommand*{\citen}[1]{%
  \begingroup
    \romannumeral-`\x 
    \setcitestyle{numbers}%
    \cite{#1}%
  \endgroup
}

\newcolumntype{.}{D{.}{.}{-1}}

\newcommand{\nbd}{\protect\nobreakdash}

\newcommand{\ra}{\rightarrow}

\newcommand{\fig}[2]{\scalebox{#1}{\includegraphics{#2}}}

\newcommand{\fxcNm}{f_{\text{xc}}^{(n-1)}}
\newcommand{\Chf}{C_{\text{HF}}}
\newcommand{\raising}[1]{\hat{a}_{#1}^\dagger}

\newcommand{\PSR}{\mathbf{P}_{\text{CO}}} 
\newcommand{\PEF}{\mathbf{P}_{\text{EF}}}
\newcommand{\FSR}{\mathbf{F}_{\text{CO}}}
\newcommand{\PSRt}{\mathbf{P}_{\text{CO}}(t')}
\newcommand{\FSRt}{\mathbf{F}_{\text{CO}}(t')}
\newcommand{\FEF}{\mathbf{F}_{\text{EF}}}
\newcommand{\Fo}{\mathbf{F}_0^{(n)}}
\newcommand{\Po}{\mathbf{P}_0^{(n)}}

\newcommand{\EATDDFT}{EA\nbd-TDDFT}
\newcommand{\EATDA}{EA\nbd-TDA}
\newcommand{\IOTDA}{IO\nbd-TDA}
\newcommand{\OODFT}{OO\nbd-DFT}

\newcommand{\revision}[1]{\textcolor{red}{#1}}

\usepackage{hyperref}
\hypersetup{colorlinks,breaklinks,linkcolor=Blue,urlcolor=Blue,anchorcolor=Blue,citecolor=Blue}
\usepackage{color}
\definecolor{DarkBlue}{rgb}{0.0,0.08,0.45}
\definecolor{Blue}{rgb}{0.0,0.0,1.0}
\definecolor{Red}{rgb}{1.0,0.0,0.0}
\definecolor{RedOrange}{rgb}{0.9,0.0,0.2}
\definecolor{dgrn}{RGB}{0,150,0}
\definecolor{dgray}{gray}{0.3}

\makeatletter
\newcommand*{\addFileDependency}[1]{
  \typeout{(#1)}
  \@addtofilelist{#1}
  \IfFileExists{#1}{}{\typeout{No file #1.}}
}
\makeatother

\newcommand*{\myexternaldocument}[1]{%
    \externaldocument{#1}%
    \addFileDependency{#1.tex}%
    \addFileDependency{#1.aux}%
}

\myexternaldocument{SI-EATDDFT}

\usepackage{titlesec}  
\titleformat{\section}[hang]{\bfseries\large}{\thesection}{1em}{}
\renewcommand{\thesection}{\arabic{section}}
\renewcommand{\thesubsection}{.\arabic{subsection}}
\renewcommand{\thesubsubsection}{.\arabic{subsubsection}}
\titleformat{\section}{\normalfont\bfseries}{\thesection.~}{0.25em}{}
\titleformat{\subsection}[runin]{\normalfont\bfseries}{\thesection\thesubsection.}{0.25em}{}
\titleformat{\subsubsection}[runin]{\normalfont\bfseries}{\thesection\thesubsection\thesubsubsection.}{0.25em}{}

\begin{document}
\title{
    Electron-Affinity Time-Dependent Density Functional Theory:
    Formalism and Applications to Core-Excited States
}

\author{
	Kevin Carter-Fenk,$^\ast$ 
	Leonardo A. Cunha,$^\ast$ 
	Juan E. Arias-Martinez,$^\ast$ 
	and Martin Head-Gordon$^\ast$
}

\affiliation{
	Kenneth S. Pitzer Center for Theoretical Chemistry, Department of Chemistry, University of California, Berkeley, CA 94720, USA\\
	Chemical Sciences Division, Lawrence Berkeley National Laboratory, Berkeley, CA 94720, USA
}
\date{\today}


\begin{abstract}\noindent 
The particle-hole interaction problem is longstanding within time-dependent density functional theory (TDDFT) and leads to extreme errors in the prediction of \revision{K-edge} X-ray absorption spectra (XAS). We derive a linear-response formalism that uses optimized orbitals of the \textit{n}--1-electron system as reference, building orbital relaxation and a proper hole into the initial density. Our approach is an exact generalization of the static-exchange approximation that ameliorates particle-hole interaction error associated with the adiabatic approximation and reduces errors in TDDFT XAS by orders of magnitude. With a statistical performance of just 0.5~eV root-mean-square error and
the same computational scaling as TDDFT under the core-valence separation approximation, we anticipate
that this approach will be of great utility in XAS calculations of large systems.
\begin{center}
TOC Graphic\\
\fig{1.0}{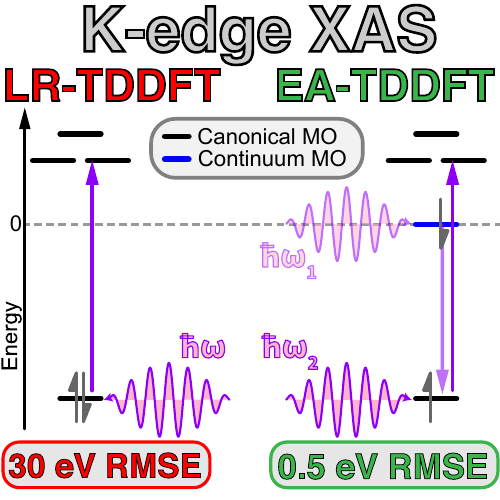}
\end{center}
\end{abstract}
\maketitle




Recent advancements in synchrotron
and ultrafast tabletop X-ray light-sources mark the dawn of an
X-ray technology renaissance.
With exceptional element-specificity,
X-ray spectroscopy has found use in probing liquid-to-metal
phase transitions of ammonia,\cite{ButMasMcM20} tracking charge-separation dynamics in dye-sensitized
solar cells and organic light-harvesting systems,\cite{GesGuh16,RotBorWen21}
and has revealed quantum nuclear dynamics near conical intersections.\cite{KatNorGaw19}
Modern X-ray absorption spectroscopy (XAS) is capable of
energy resolution on the order of 0.2--0.9~eV,\cite{ZimPerAbd20}
which is well below the error statistics of most modern theoretical methods that are routinely used to model XAS. 

Linear-response time-dependent density functional theory (TDDFT) 
is by far the most commonly used
method for computing excitation energies
due its accuracy and efficiency.\cite{Cas95a,PetGosGro96,BauAhl96a,Fur01a,MarGro04,MarMaiNog12}
While formally exact for excitation energies, TDDFT in practice
is approximate due to inexact ground state functionals
and the ubiquitous adiabatic approximation (henceforth assumed).\cite{Cas95a,CasHui12,BurWerGro05}
Although TDDFT achieves statistical accuracy of 
$\sim$0.2--0.3~eV
for valence excitations,\cite{LauJac13}
errors increase dramatically for core excitations,
often requiring empirical shifts on the order of 10--100~eV to realign the calculated
spectra with experiment.\cite{LesNguLi15,ChaKowMag18,BusHut21a,BusHut21b}
\revision{While range-separated hybrid functionals perform
better in this regard,}\cite{doCHolSla15,FraBruVid21}
specialized short-range corrected (SRC) functionals
that feature a large amount of short-range Hartree-Fock (HF) exchange
to correct for differential self-interaction error
in the core
have also been used instead of empirical shifting,
albeit to the disregard of broader thermochemical
properties.\cite{SonWatNak08,SonWatHir09,BesPeaToz09,BesAsm10,CapPenBes13,Bes16,Bes20,Bes21}
Apart from pure TDDFT,
semi-empirical extensions of configuration interaction
that employ Kohn-Sham orbitals
have been applied with some success to core-excitations.\cite{RoeMagDeB13,MagRoeHav13,SeiNevKle19}
\revision{In some cases, particularly in periodic systems,
TDDFT and configuration-interaction methods are sidestepped
in favor of cruder approaches like the Slater transition
or transition potential
methods.\cite{SlaWoo71,Sla72,SteLisDec95,TriPetAgr98a,TriPetAgr98b,CavOdeNor05,LeeLjuLyu10,FraZhoCor16,MicReu19}
Despite the myriad ways in which XAS can be calculated,
in this work our focal point is linear-response TDDFT.
}

One source of error in predicting core-excitations
using linear-response theory is the
large orbital relaxation effect that follows from the
displacement of charge out of a core orbital.\cite{JanBanMuk99,RanPen21}
This can be addressed on a state-by-state basis
using orbital-optimized density functional theory (\OODFT), which
explicitly relaxes the orbitals of excited-state configurations.\cite{HaiHea21}
While \OODFT\ routinely achieves a statistical accuracy of $\sim$0.3~eV
for core-excitations,\cite{HaiHea20b,CunHaiKan22}
state-by-state optimization is far less efficient
than full-spectrum methods like TDDFT.
\OODFT\ also requires some {\em a priori} knowledge of the system,
complicating the selection of the
``correct'' set of bespoke determinants
in systems with a high density of states.

Orbital relaxation error is related to 
the fundamentally
incorrect particle-hole interaction in
TDDFT descriptions of core-excited states.\cite{DreHea04,DreHea05,Mai05}
This is the major source of error in TDDFT; emerging from the fact that
the virtual orbitals in DFT are optimized in the $n$-electron potential,
causing incomplete cancellation of the interaction of 
the excited electron with itself
in the (previously occupied) core orbital.
For example, consider the pure particle-hole interaction
that results from exciting an electron
between two molecular orbitals (MOs)
that have zero overlap.
For global hybrid functionals, 
the only nonzero
elements of the orbital Hessians belong to
the $\mathbf{A}$ 
matrix,
\begin{equation}\label{eq:Example}
        A_{ia,jb} = (\varepsilon_a - \varepsilon_i)\delta_{ij}\delta_{ab}
        - \Chf(ij|ab)
\end{equation}
where $\Chf$ is the coefficient of HF exchange.
The Coulomb interaction $(aa|ii)$ is included in the orbital energy difference
and only in the case of exact exchange ($\Chf=1$) is
this interaction properly cancelled by the third term in Eq.~\ref{eq:Example},
leading to
particle-hole attraction.
Therefore with approximate density functionals,
the excited electron ``feels'' a residual Coulomb potential from
its unexcited image rather than a proper particle-hole attraction,
causing core-excitation energies
to be dramatically underestimated in approximate
TDDFT.

In this letter, we introduce a linear-response TDDFT formalism
that effectively models particle creation in the virtual space from an
\textit{n}--1-electron reference density.
This way, orbital relaxation
and information about the core hole are built directly
into the reference density, completely eliminating
electron-hole self-interaction error (eh-SIE) by construction.
Our method generalizes the static-exchange approximation (STEX)
into a density functional theory (DFT) framework.

Such generalizations have recently been proposed based purely on error
cancellation between restricted open-shell Kohn-Sham (ROKS) theory
and STEX,\cite{HaiOosCar22} but this work aims at a fully derivable formalism.
Herein we demonstrate multifaceted
benefits of an exact approach,
including 
better overall performance,
and a recovery of the
Jacob's Ladder concept in DFT.
While TDDFT is the workhorse of excited-state calculations in
quantum chemistry, we further note that the concept of adding electrons to
an \textit{n}--1-electron reference determinant has been employed
within Green's function based GW methods\cite{DucDeuBla12,JinYan19}
and within algebraic diagrammatic construction approaches\cite{FraDre19,DreFra22}
to account for orbital relaxation
and (in the case of GW theory) for a more appropriate
description of particle-hole interactions.
In principle, our proposed approach recovers the same poles as
the single-particle Green's function in the electron addition domain.
However, unlike GW approaches that scale roughly as
${\mathcal O}(N^4)$ with a nontrivial prefactor (where $N$ is the
number of basis functions),\cite{WilSeeGol21}
our proposed approach has the same scaling as STEX
(${\mathcal O}[V^3]$, where $V$ is the number of
unoccupied MOs), making this a far more appealing method for large systems.

The STEX formalism has been used to improve upon core-excitation energies
offered by configuration interaction with single excitations (CIS)\cite{DelDitPop71,ForHeaPop92,DreHea05}
for a number of years.\cite{AgrCarVah94,AgrCarVah97}
In brief, STEX involves optimizing the MOs of the \textit{n}--1-electron
(core-ionized) system, followed by an electron-affinity CIS (EA-CIS)
calculation to reattach the missing electron to the virtual orbitals,
thereby yielding a partially orbital-optimized
core-excitation spectrum that accounts for the strong polarization
effect from creating a core hole.
For a closed-shell reference, the EA-CIS equations
for singlet 
states take the form,
\begin{equation}\label{eq:EACIS}
    A_{ia,ib} = E^{(n-1)}_{\text{HF}}\delta_{ab}+F_{ab}^{(n-1)} + (ia|ib) \;,
\end{equation}
where $i$ is the core hole MO,
$E_{\text{HF}}^{(n-1)}$ is the core-ionized reference energy,
$F_{ab}^{(n-1)}$ are elements of the virtual-virtual block of
the core-ionized Fock matrix, and $(ia|ib)$ is an exchange integral
in the standard Mulliken notation.
Diagonalizing $\mathbf{A}$ results in states that are orthogonal to
the core-ionized reference determinant, but are not orthogonal to
the original $n$-electron ground state.
The final step of the STEX procedure involves constructing
nonorthogonal configuration interaction (NOCI) elements
to project the $n$-electron ground state out of the
Hamiltonian prior to diagonalizing, ensuring that all
excited states are strictly orthogonal to the
initial $n$-electron ground state determinant.\cite{OosWhiHea18,OosWhiHea19,OosWhiHai20}
\revision{Herein, we will show that the nonorthogonality of
the excited states
to the ground state can be safely
ignored when calculating K-edge XAS
with almost no impact on the predicted excitation energies
or transition properties,}
thus paving the way
for a TDDFT formalism where the ambiguity of
DFT-based NOCI elements
once hindered such developments.

In order to generalize 
STEX 
to a TDDFT framework,
we will use continuum MOs as a derivation tool.
For our purposes, continuum MOs are fictitious, ultra-diffuse orbitals that do not interact
with other MOs in the system and have zero energy.
They offer utility in derivations of 
particle-nonconserving processes
by recasting particle creation\slash annihilation into
the language of particle-conserving excitations.\cite{LiuHatHof20}
Throughout 
this work,
we reserve the labels
\{$j,k,l,$\dots\} to denote occupied MOs, \{$a,b,c,$\dots\}
for the virtual MOs, and \{$p,q,r,$\dots\} refer to general orbitals.
Specific notation is reserved for the
continuum MO, designated as $x$,
and the core-hole MO, $i$.

To ameliorate \revision{orbital relaxation error,} 
we begin with the self-consistently optimized MOs for
the core-ionized system.
We are interested in a protocol that uses particle-conserving
excitations that emulate the action of the particle creation
operator on our core-ionized reference,
\begin{equation}\label{eq:creation}
    |\Psi_i^a\rangle = \raising{a}|\Psi_0\rangle \;,
\end{equation}
where $|\Psi_0\rangle$ is the core-ionized reference determinant,
and $\raising{a}$ is the creation operator.
One possibility that retains correlations between single excitations
in the response theory that follows is to consider two
successive excitations $x\rightarrow i$ and $i \rightarrow a$
out of a modified core-ionized reference determinant that includes
a single continuum MO.
Conceptually, this can be likened to
excited-state absorption where the $n$-electron
state with the core-hole MO reoccupied acts as the intermediate state.
In operator form it can be readily shown that,
\begin{equation}\label{eq:wicks}
    |\Psi_i^a\rangle =
    \raising{a}\hat{a}_i^{\vphantom{\dagger}}{\hat{a}_i}^\dagger
    \hat{a}_x^{\vphantom{\dagger}}|\Psi_0\chi_x\rangle
    =
    \raising{a}\hat{a}_i^{\vphantom{\dagger}}\hat{a}_i^\dagger
    |\Psi_0\rangle
    = \raising{a}|\Psi_0\rangle \;,
\end{equation}
where $|\Psi_0\chi_x \rangle$ is the modified core-ionized reference,
containing the noninteracting spin-orbital $\chi_x$.
This exercise reveals that
the successive particle-conserving excitations
$x\rightarrow i$ and $i \rightarrow a$ indeed reduce to
particle creation in orbital $a$ of the
{\em unmodified} core-ionized reference determinant,
which itself can be viewed as the tensor product of the
core-ionized reference with the vacuum level in the space of
continuum orbitals.

In order to capture this process in the language of
density matrices,
such that our protocol is amenable to
DFT,
we \revision{consider} two successive linear responses.
The first response generates the $n$-electron density from
the \textit{n}--1-electron reference by exciting an electron from a
continuum MO into the core hole, and the
second response
\revision{yields eh-SIE-corrected
excitations of this (newly added)
core electron into the virtual space.}
Throughout this derivation, we follow the
density matrix formalism starting from the
Liouville-von Neumann equation,\cite{Fur01a,DreHea05}
\begin{equation}\label{eq:LvN}
    i\frac{\partial\mathbf{P}(t)}{\partial t} = [\mathbf{F}(t),\mathbf{P}(t)] \;.
\end{equation}

The first response 
is obtained by restricting
the excitation space to the (occupied) continuum MO
and the (unoccupied) core-hole to yield,
\begin{equation}\label{eq:continuumResp}
    \begin{split}
        A_{xi,xi} &= F_{ii}^{(n-1)}\\
        B_{xi,xi} &= 0
    \end{split}
\end{equation}
\revision{Because the continuum MO does not interact with the rest of
the system,}
all two-electron integrals involving the continuum MO vanish
to give an expression that corresponds to the
negative electron affinity in the limit of the exact functional.\cite{TsuSonSuz10}
Importantly, no orbital rotations are encoded in this response,
meaning that the (idempotent) $n$-electron density
can be exactly constructed with the \textit{n}--1-electron MOs to linear order.


At some time $t'>t$, we apply a second time-varying electric field
to the perturbed $n$-electron system.
By regenerating the $n$-electron system, we have reintroduced
eh-SIE, so we now seek
to separate the response due to the presence of the core electron
from the remainder of the response to the applied field.
Assuming that the 
$n$-electron density is not too
far from a stationary point, we may write the perturbed density
and corresponding Fock matrices as,
\begin{equation}\label{eq:DoubleResp}
    \begin{split}
        \mathbf{P}(t') &= \mathbf{P}^{(n)}_0 + \delta\PSR(t') + \delta\PEF(t')\\
        \mathbf{F}(t') &= \mathbf{F}^{(n)}_0 + \delta\FSR(t') + \delta\FEF(t')
    \end{split}
\end{equation}
where $\mathbf{P}^{(n)}_0$ is the static part of the $n$-electron density,
$\delta\PSR(t')$ represents the component of the response due to
the (now occupied) core MO
and $\delta\PEF(t')$ indicates the response of the $n$-electron system
to the second electric field.
Substituting Eq.~\ref{eq:DoubleResp} into Eq.~\ref{eq:LvN}
and keeping the terms that are linear with respect to the perturbing field
leads to,
\begin{equation}\label{eq:DLR}
    \begin{split}
	&i\frac{\partial\delta\PSRt}{\partial t'}+i\frac{\partial\delta\mathbf{P}_{\rm EF}(t')}{\partial t'} =
	[\Fo,\delta\PSRt]\\
	&+[\delta\FSRt,\Po]
	+[\Fo,\delta\mathbf{P}_{\rm EF}(t')]+[\delta\mathbf{F}_{\rm EF}(t'),\Po]
    \end{split} \;,
\end{equation}
which is simply the sum of two linear responses.
Notably, this formalism has been used to subtract nonstationary oscillations
out of real-time TDDFT
simulations of excited-state absorption \revision{(including application to transient XAS)},
and we have adopted similar
notation throughout.\cite{FisCraGov15,BowAshFis17,CavNasZha21,LieHoWea21,LoeLieGov21}

We use the fact that the above responses are uncoupled
to correct the
$n$-electron response
by subtracting the
components
that emerge due to the occupied
core orbital {\em via} the difference Fock matrix,
\begin{equation}\label{eq:NSFock}
\begin{split}
    F^{\text{CO}}_{pq} &= F_{pq}^{(n)} - F_{pq}^{(n-1)}\\
    &= (ii|pq)-\Chf(ip|iq)\\
    &\qquad+(1-\Chf)(p|V_{\text{xc}}^{(n)}-V_{\text{xc}}^{(n-1)}|q)
\end{split}
\end{equation}
and its corresponding density 
(all $n$-electron quantities are constructed using the \textit{n}--1-electron MOs).
This form of the Fock matrix incorporates all zeroth-order couplings
between $n$- and \textit{n}--1-electron potentials
without approximation, and the corresponding difference density
is idempotent with one electron in core MO $i$, permitting excitations
of the form $i\ra a$.

Subtracting the response of the
core orbital
density from
that of the $n$-electron density (see Sec.~\ref{sec:Partials} for details),
leads to a
\revision{eh-SIE-}corrected $n$-electron response in terms of \textit{n}--1-electron quantities
\begin{equation}\label{eq:EATDDFT_2}
    \begin{split}
        A_{ia,ib} &= F_{ab}^{(n-1)} - F_{ii}^{(n-1)}\delta_{ab}\\
        &\qquad + (ia|ib) + (1-\Chf)(ia|\fxcNm|ib)\\
        B_{ia,ib} &= (ia|ib) + (1-\Chf)(ia|\fxcNm|ib)
    \end{split}
\end{equation}
where,
\begin{equation}
\fxcNm = \frac{\partial V_{\text{xc}}[\rho^{(n-1)}]}{\partial\rho^{(n-1)}} \;.
\end{equation}
Finally, we add this to the result of the initial response in Eq.~\ref{eq:continuumResp}
to obtain the working equations,
\begin{equation}\label{eq:EATDDFT}
    \begin{split}
        A_{ia,ib} &= F_{ab}^{(n-1)} + (ia|ib) + (1-\Chf)(ia|\fxcNm|ib)\\
        B_{ia,ib} &= (ia|ib) + (1-\Chf)(ia|\fxcNm|ib)
    \end{split}
\end{equation}

Each of the above responses comes from exact TDDFT, suggesting
that with time-dependent exchange-correlation kernels this approach could be
made exact (to first order).
Of course, knowledge of the exact functional would render
this formalism obsolete
because the exact functional is asymptotically
correct (eh-SIE-free) and has
frequency dependence (accounting for orbital relaxation).\cite{Mai17}
From a utilitarian perspective, the exact functional is not available
and all practical TDDFT implementations employ the adiabatic
local density approximation (ALDA).
Therefore, Eq.~\ref{eq:EATDDFT} can be viewed as a pragmatic
correction to errors associated with the ALDA in TDDFT for XAS.

By nature of the core-ionized reference determinant
and because the MOs do not relax
on addition of the electron,
the orbital relaxation codified into the \textit{n}--1-electron density
is retained.
\revision{The second response is also eh-SIE-corrected,
ensuring that there is no residual Coulomb-like interaction
between excited electron and core hole.
In fact, our proposed correction (Eq.~\ref{eq:NSResponse_SI})
bears a delightful
resemblance to
the virtual-orbital self-interaction definition proposed by
Imamura and Nakai,\cite{ImaNak07}
with the added benefit that our equations capture
orbital relaxation.}
This immediately suggests a metric for quantifying the
extent of eh-SIE
{\em via} the eigenvalues of the core-orbital response matrix (Eq.~\ref{eq:metric}).
In the limit of the Hartree-Fock functional ($\Chf=1$)
this metric is exactly zero,
and under the Tamm-Dancoff approximation (TDA)\cite{HirHea99b}
Eq.~\ref{eq:EATDDFT} becomes precisely equivalent
to the EA-CIS equation (Eq.~\ref{eq:EACIS}).
This
implies that Eq.~\ref{eq:EATDDFT} is a generalization
of EA-CIS to a DFT framework,
so we call our approach
electron-affinity TDDFT (\EATDDFT).


In true analogy to EA-CIS,
we will employ the TDA (\EATDA) throughout this work,
setting the $\mathbf{B}$ matrix to zero in Eq.~\ref{eq:EATDDFT}.
\revision{
While in some applications the modified reference
state ({\em e.g.} a core-ionized determinant)
can lead to difficulties in solving the full
non-Hermitian eigenvalue problem due to orbital
rotations that drive the solution towards the ground
state,\cite{GosWei80,WeiGos80,PraPalMuk85,DatMukMuk93}
we have found that such problems are not encountered in \EATDDFT.\cite{CarHea22}
Despite the fact that the full \EATDDFT\ equations
can be readily solved, the TDA is likely
an excellent approximation within the confines of
core excitations associated with K-edge XAS,
as the elements of $\mathbf{B}$ are quite small.
For the sake of comparison,
\EATDA\ is also a more direct analogue to STEX (a CIS theory),
which is what we are attempting to generalize to DFT.
}

\begin{figure}[ht!!]
\centering
\fig{1.0}{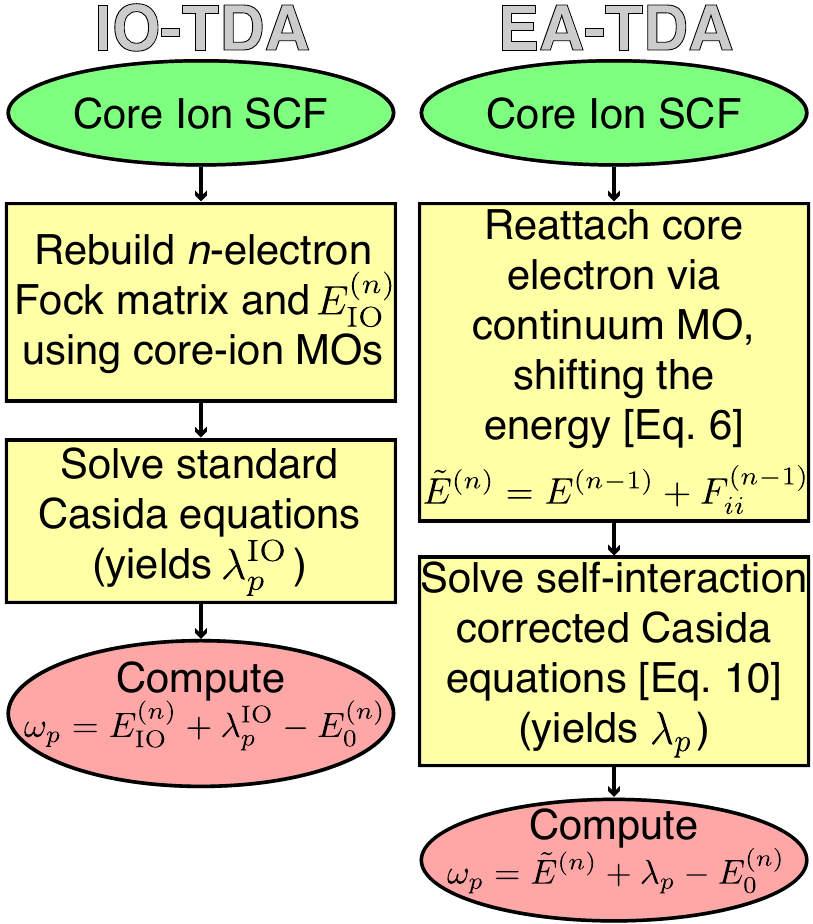}
\caption{
Flowchart describing the conceptual protocol for (left) an
\IOTDA\ calculation and (right) an \EATDA\ calculation,
where $E_{\text{IO}}^{(n)}$ is the energy of
the \textit{n}-electron Fock matrix
constructed {\em via} core-ion orbitals,
$E_0^{(n)}$ is the SCF ground-state energy,
$\lambda_p^{\text{IO}}$ are the eigenvalues of
the Casida equations using the \IOTDA\ reference,
$\lambda_p$ are the eh-SIE corrected eigenvalues
of \EATDA, and $\omega_p$ are the core-excitation
energies.
}
\label{fig:Flowchart}
\end{figure}

Apart from \EATDA,
we 
also consider the more naive approach
of optimizing the orbitals of the \textit{n}--1-electron system
and using them directly to reconstruct the $n$-electron
density. From this nonstationary initial state, we
perform TDDFT under the TDA using the usual Casida formulation.
In principle, this ion-orbital TDA (\IOTDA) approach
incorporates orbital relaxation into the reference
but lacks the ingredients that account for
eh-SIE
(details in Section~\ref{sec:OOTDDFTSI}).

\revision{
It is important to note that \EATDA\ differs
strongly from \IOTDA\ in two respects.
First, the energy of the intermediate $n$-electron
state is constructed differently between the two methods.
Whereas \IOTDA\ forms $E_{\text{IO}}^{(n)}$,
the energy of the \textit{n}-electron Fock matrix
using the orbitals of the core-ion with
the core electron reattached,
\EATDA\ constructs $\tilde{E}^{(n)}=E^{(n-1)}+F_{ii}^{(n-1)}$.
The two energies are only
equivalent in the case of HF
where the Fock matrix is strictly linear
in the density.
Second, we emphasize that Eq.~\ref{eq:EATDDFT_2} is
not just a standard linear-response TDDFT expression, but
one that encodes an exact first-order eh-SIE
correction to the TDDFT equations.
Without this correction, \EATDA\ would correct for orbital
relaxation but would not correct for the (even larger)
self-interaction error in the excited state.
The flowchart in Fig.~\ref{fig:Flowchart}
emphasizes the difference between the
eigenvalues $\lambda_p^{\text{IO}}$ of \IOTDA,
which corrects for orbital relaxation but not eh-SIE,
and the self-interaction corrected eigenvalues, $\lambda_p$, of \EATDA.
Later in this work, we investigate the impact of
orbital relaxation and eh-SIE, showing that both must be adequately
compensated for accurate results.
Finally, we note that explicitly performing each step in the
conceptual protocol in Fig.~\ref{fig:Flowchart}
is not necessary in practice, where Eq.~\ref{eq:EATDDFT}
can be directly constructed and diagonalized to yield
electron affinities $\gamma_p$ such that
$\omega_p=E^{(n-1)}+\gamma_p - E_0^{(n)}$.
}

\begin{figure*}[ht!!]
    \centering
    \fig{1.0}{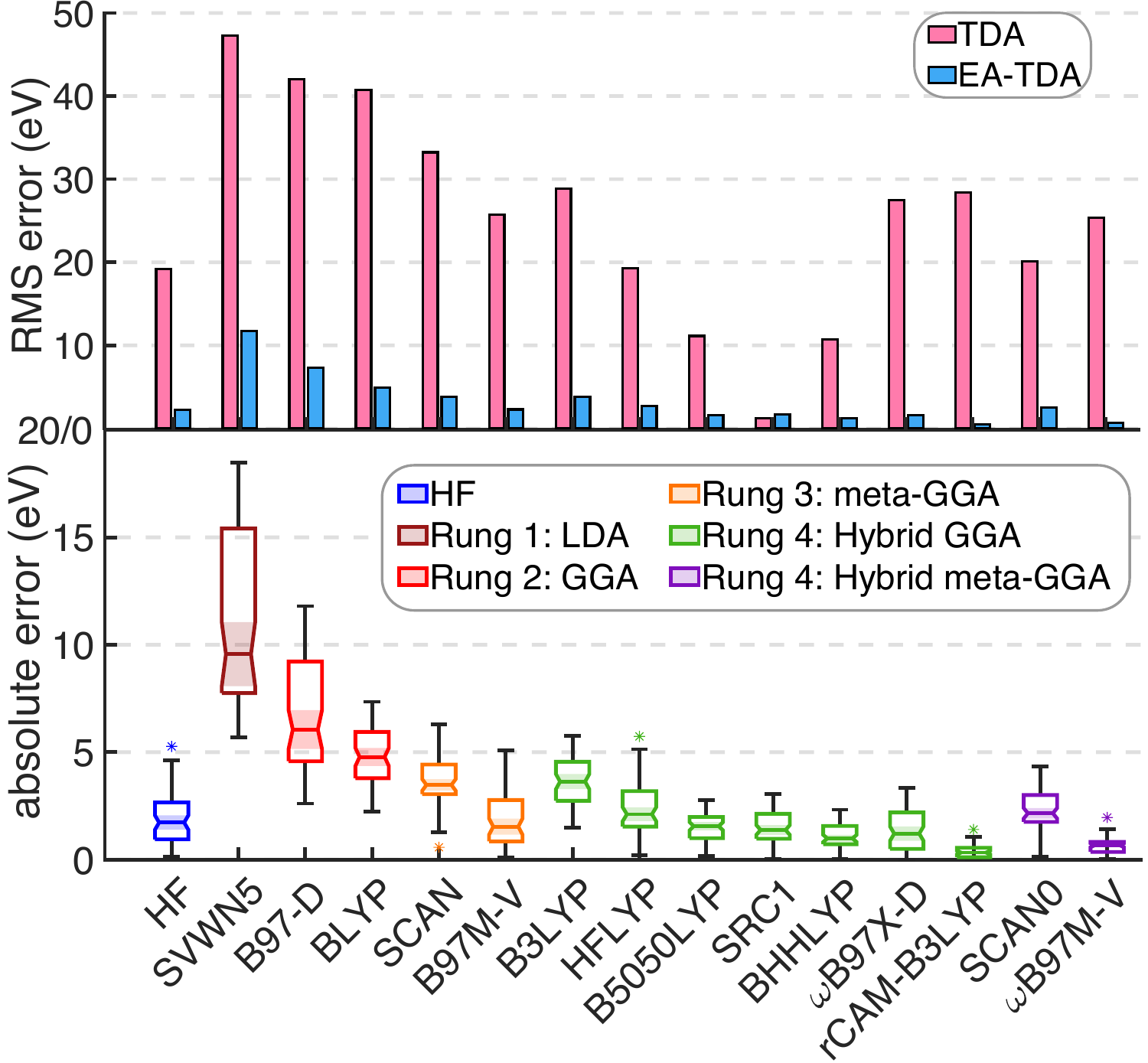}
    \caption{Absolute error statistics for 65 experimental K-edge transitions
    (lowest energy transition only).
    (Top) RMSE for standard TDA versus \EATDA\ by functional.
    (Bottom)
    \EATDA\ absolute error statistics.
    Upper and lower delimiters indicate maximum
    and minimum errors, respectively. Upper and
    lower bounds of each box are the upper and lower
    quartiles, respectively. Median absolute errors
    are indicated by horizontal lines and
    overlapping notches
    identify statistical similarities between
    distributions to the 95\% confidence level.
    Outliers are indicated by asterisks.
    All calculations use aug-pcseg-1 for H and Br atoms
    and aug-pcX-2 otherwise.
    }
    \label{fig:JacobsLadder}
\end{figure*}

\revision{This definition of the excitation energies
is common between \EATDA\ and STEX,
but unlike STEX, \EATDA\ does not use a projection operator to ensure
orthogonality between the singly-excited configurations and
the $n$-electron ground state.}
Nonorthogonality can have detrimental effects on transition dipole moments (TDMs)
even when energies are largely unaffected,\cite{WorFeiMan21}
so we take it into account by using a pseudo-wavefunction
{\em ansatz} (as done in TDDFT) to compute an overlap-free TDM
(see Sec.~\ref{sec:TDMs} for details).
We justify this approach by comparison of \EATDA\ with the HF
functional against STEX
on a data set of 132 experimental K-edge transitions of small molecules.
Tables \ref{table:EATDAvSTEX_1}--\ref{table:EATDAvSTEX_3}
reveal strong agreement between \EATDA(HF) and STEX,
with a maximum difference in transition energies
of just 0.1~eV and a mean difference of 0.01~eV.
The average difference between \EATDA(HF) and STEX oscillator strengths is
$\sim 10^{-5}$, with a maximum difference of $\sim 10^{-4}$
while the average STEX oscillator strength is $10^{-2}$.
We therefore conclude that the
nonorthogonality of the final states in \EATDA\
exerts a minimal influence on the details of the spectrum.


We assess the functional dependence of \EATDA\ across
15 density functionals, and while not comprehensive
we include data from local density approximation (LDA),
generalized gradient approximation (GGA),
meta-GGA, hybrid GGA, and hybrid meta-GGA functionals.
The bottom panel of Fig.~\ref{fig:JacobsLadder} indicates
a clean recovery of the Jacob's Ladder concept in
DFT, with errors decreasing with each step up through the rungs.
Signed errors (Fig.~\ref{fig:MSE}) reveal that
semi-local
functionals tend to underestimate
excitation energies, whereas asymptotically correct functionals
exhibit very little systematic error.
Increasing the fraction of exact exchange improves
error statistics up to a point with root mean squared error (RMSE)
decreasing from BLYP to B3LYP and on to B5050LYP,
but too much exact exchange
degrades the results leading to higher RMSE for HFLYP than B5050LYP.
Overall, asymptotically correct functionals perform best,
and among them rCAM-B3LYP performs best of all
with an RMSE of only 0.5~eV.

To understand the scope of the improvements
offered by \EATDA, 
we compare RMSEs of standard TDA with \EATDA\
across functionals.
The top panel of Fig.~\ref{fig:JacobsLadder}
reveals that for a given functional
the average improvement offered by \EATDA\ is on
the order of tens of eV.
In fact, the RMSE of rCAM-B3LYP improves by
roughly two orders of magnitude, from 28.4~eV to
0.5~eV.
This massive improvement is apparent in all but
the SRC1 functional, which was parameterized
specifically to cancel eh-SIE
in standard TDDFT. This is a testament to the parameterization
of SRC1, but the lack of improvement (or deterioration) of the results
on switching to \EATDA\
also exemplifies that eh-SIE 
is adequately quenched in \EATDA.

\begin{figure}
\centering
\fig{1.0}{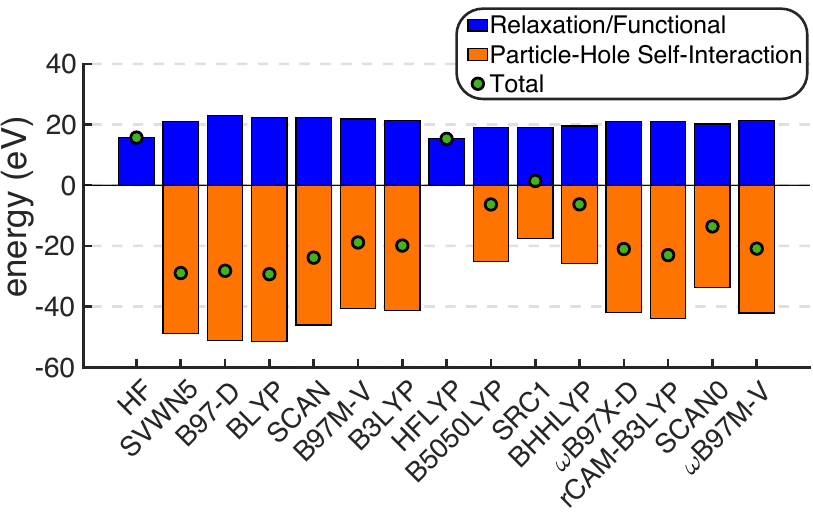}
\caption{
Error contributions to standard TDA that are corrected in
\EATDA, averaged over 65 K-edge transitions.
The total (signed) error is taken to be the difference
$Err(\text{TDA}) - Err(\text{\EATDA})$,
which is equivalent to the sum of the bars.
}\label{fig:ErrorDecomp}
\end{figure}

On 
examination of Eq.~\ref{eq:EATDDFT} it becomes clear
that a necessary criterion for a functional to perform
well with \EATDA\ is an accurate estimate of the electron affinity
for each virtual MO. This is because the dominant term is
$F_{ab}^{(n-1)}$, while the last two terms offer only small corrections
to this energy because they are dependent on the overlap
of the core orbital with the virtual MOs.
Because asymptotically correct functionals perform best
in the prediction of electron affinities,\cite{AndOviWon17,ZhoHuJia18}
so too do they perform best with \EATDA.

The success of \EATDA, a theory that takes orbital relaxation
and eh-SIE into account, allows us to diagnose the origins of
the errors in standard TDA. Using the metric defined in
\ref{eq:metric} we have immediate access to the amount of eh-SIE
in standard TDA, while the remainder of the error in TDA
can be ascribed to orbital relaxation effects.
In Fig.~\ref{fig:ErrorDecomp}, we
define the total error (relaxation error plus eh-SIE)
as the difference between standard TDA and \EATDA,
$Err(\text{TDA}) - Err(\text{\EATDA})$, where
the errors in excitation energies $\omega_{\text{X}}$
of a given method $\text{X}$ are taken
to be $Err(\text{X})=\omega_{\text{X}}-\omega_{\text{ref}}$
with $\omega_{\text{ref}}$ representing the experimental value.
Orbital relaxation contributes positive errors
because without relaxation effects the predicted excitation energies are higher,
whereas eh-SIE over-stabilizes the excitation energy due to
a lack of particle-hole attraction, so its contribution is
net negative.

The HF functional has zero contribution from eh-SIE because
exact exchange yields correct particle-hole attraction.
Similarly, HFLYP has a near-zero contribution from eh-SIE due to exact
exchange, but it does not substantially improve upon HF in terms of orbital relaxation.
Otherwise, the lion's share of error in most functionals comes from
eh-SIE, with a relatively consistent contribution from a lack of orbital
relaxation.
An interesting exception to this rule is the SRC1 functional,
which (owing to its parameterization for XAS) hosts
a nearly optimal degree of error cancellation between orbital relaxation error
and eh-SIE at a respective ratio of $52:48$.
Other functionals that benefit from near-cancellation of
errors are those with a large fraction of global HF exchange
such as B5050LYP and BHHLYP, which explains their notably better performance
in comparison to other functionals in the top panel of Fig.~\ref{fig:JacobsLadder}.
Overall, the largest contribution to the errors in TDA
are from eh-SIE while orbital relaxation plays a consistent,
strong, but auxiliary role in defining the total error.

\begin{figure}
    \centering
    \fig{1.0}{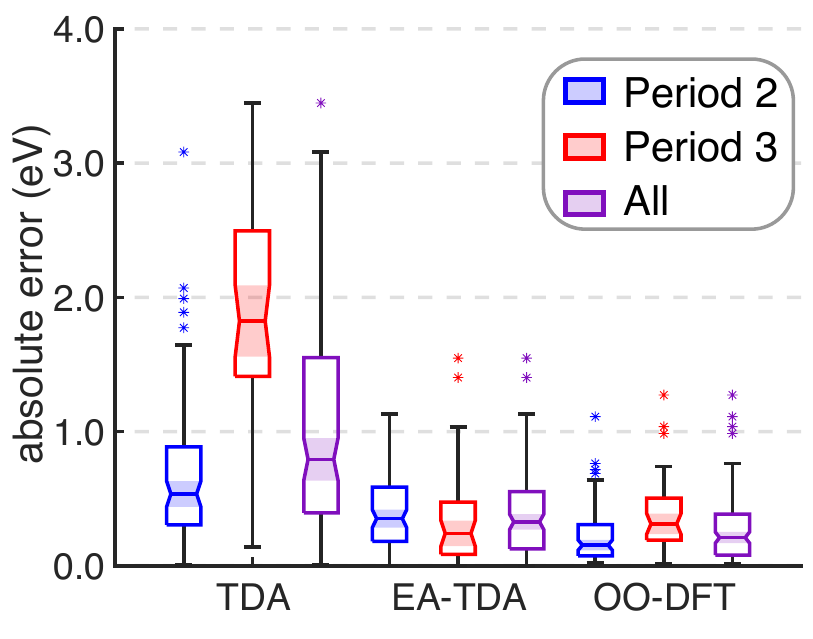}
    \caption{Absolute error statistics for
    the best method\slash functional combinations
    against 132 experimental K-edge transitions.
    Results are broken down by Period alongside
    the full data set. For TDA, SRC1-RX was
    used (where X = 1 or 2, depending on the period),
    rCAM-B3LYP and SCAN were used for \EATDA\
    and \OODFT, respectively.
    All calculations use aug-pcseg-1 for H and Br
    and the doubly-augmented
    d-aug-pcX-2 basis otherwise.
    }
    \label{fig:Best}
\end{figure}

We repeated
the statistical analysis in Fig.~\ref{fig:JacobsLadder} for \IOTDA\
and standard TDA to find the best functionals
for each method (see Fig.~\ref{fig:OOTDAerrors}
and Fig.~\ref{fig:LRTDAerrors} for detailed statistics).
Our group has previously established SCAN as an excellent
functional for core-excitations with \OODFT, so we forego
further analysis here.\cite{HaiHea20b,CunHaiKan22}
The method\slash functional combinations that
yielded the lowest errors on this test set were
subjected to the more rigorous test of 132 experimental
K-edge transitions of 46 molecules,
ranging from 1--5 transitions per molecule.
The results for \IOTDA\ were omitted because
the best functional for \IOTDA\ was HF,
suggesting that DFT provides no benefit to nonstationary
TDA if eh-SIE is not properly taken into account.
The results in Fig.~\ref{fig:Best}
suggest that \EATDA\ ($\rm RMSE = 0.5$~eV) is almost as accurate as
\OODFT\ ($\rm RMSE = 0.4$~eV) across the board,
outperforming the SRC1 functionals
used with standard TDA ($\rm RMSE = 1.3$~eV)
even though SRC1 was
designed to accurately capture core-excitation energies.

\begin{figure}
    \centering
    \fig{1.0}{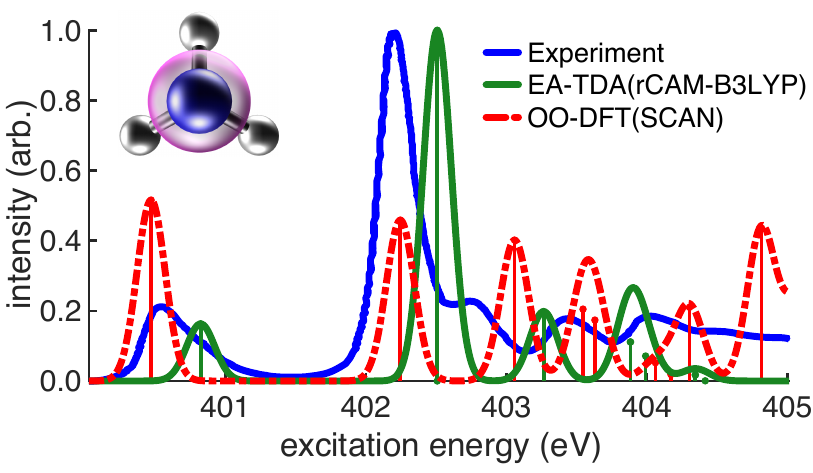}
    \caption{
    Ammonia K-edge X-ray absorption spectra
    for \EATDA\ and \OODFT\ juxtaposed
    against experimental data from
    Ref.~\citen{SchTroRan93}.
    The aug-pcX-2 and aug-pcseg-1 basis sets
    were used for N and H, respectively.
    }
    \label{fig:FullSpectra}
\end{figure}

\EATDA\ also has excellent performance with respect to
atomic size. While TDA and \OODFT\ results show
a statistically significant increase in errors from
Period 2 elements to Period 3
(indicated by nonoverlapping notches in Fig.~\ref{fig:Best}),
\EATDA\ results remain statistically similar
with an RMSE that is equivalent to 
that of \OODFT\ (0.5~eV).
Admittedly, \EATDA\ exhibits a slightly wider
distribution of errors than \OODFT, as shown by the quartiles.
Being that the entire spectrum is obtained with one single
\EATDA\ calculation whereas \OODFT\ optimizes specific
configurations, this level of comparability between
results is quite satisfactory.

The excellent comparability of \EATDA\ and \OODFT\
excitation energies
begs the question: do 
transition
properties also behave similarly?
To investigate this,
the experimental K-edge spectrum of ammonia 
is shown in Fig.~\ref{fig:FullSpectra} alongside
calculated spectra using \EATDA\ and \OODFT.
For this system, \OODFT\ predicts slightly better
excitation energies, whereas \EATDA\
shows a slight
(0.2~eV) blue shift.
Insofar as transition strengths are concerned,
the first major peak
in the \OODFT\ spectrum (normalized intensity of 0.52) corresponds
to the 1s$\ra$3s transition, which is largely symmetry forbidden,
leading to a small pre-edge peak in the experiment.
The amplification of this feature
is an artifact of ROKS, which allows
strong singlet-triplet mixing
when the initial and final states 
are isosymmetric.\cite{GriNonFra03,BilEgl06,FilSha99b,FriDamFra08,KowTsuChe13}
Interestingly, the errors do
not seem to stem from nonorthogonality of the ROKS states,
as we account for residual nonorthogonality
by subtracting the overlap-weighted nuclear dipole moment.\cite{WorFeiMan21}
Instead, these errors may emerge as a direct consequence of spurious spatial symmetry-breaking due to use of approximate functionals.\cite{Bar79,Gor93,LevNag99}
Spuriously large oscillator strengths occur frequently for 1s$\ra$3s transitions
in ROKS spectra,
exerting a catastrophic impact on spectra containing truly dark 
1s$\ra$3s transitions like the one in
{\em trans}-butadiene (Fig.~\ref{fig:butadiene}).
The \EATDA\ spectrum is devoid of such spurious
high-intensity dark states and the overall
qualitative nature of the spectrum is captured
to high fidelity in all cases.

Overall, we find that \EATDA\ successfully ameliorates eh-SIE,
providing sizable improvements over TDDFT-based response theory,
yielding core-excitation energies on par with \OODFT\
while avoiding the spurious high-intensity dark states that occur in the latter.
We anticipate that \EATDA, with its low computational cost
and high overall accuracy, will be a tool of great importance in
condensed-phase systems.
Our group is currently assessing the possibility of
applying \EATDA\ to liquid-phase XAS.\cite{CarHea22}


\section*{Computational Details}
All calculations were performed with a development version of
the Q-Chem~5.4 software package.\cite{QCHEM5}
The DFT calculations use a dense quadrature with 99 radial
and 590 angular grid points to evaluate the exchange-correlation
potential, and scalar relativistic effects are accounted
for using a spin-free exact two-component
(X2C) model.\cite{VerDerEva16,Dya97,KutLiu05,IliSau07,LiuPen09,Sau11,LiXiaLiu12,CheGau11,CunHaiKan22}
Core-ionized references were optimized using
restricted open-shell orbitals and the solutions
were stabilized using combinations of
maximum overlap method (MOM),\cite{GilBesGil08,BarGilGil18b}
square-gradient minimization (SGM),\cite{HaiHea20a}
and state-targeted energy projection (STEP).\cite{CarHer20b}
We use experimental molecular geometries whenever possible
and all geometries are available in the Supporting Information.

\section*{Supporting Information}
Additional details pertaining to the derivation presented
including a description of \IOTDA,
raw data comparing \EATDA(HF) with STEX,
signed error statistics of \EATDA,
absolute error statistics for \IOTDA\ and TDA,
and spectra of {\em trans}-butadiene (PDF).
All geometries used for K-edge XAS calculations (TXT).

\section*{Acknowledgements}
This work was supported by the Director, Office of Science, Office of Basic Energy Sciences, of the U.S. Department of Energy under Contract No. DE-AC02-05CH11231.

%
%

\providecommand{\refin}[1]{\\ \textbf{Referenced in:} #1}

\makeatletter\@input{xx.tex}\makeatother

\end{document}


\title{
	Supporting Information for:\\
``Electron-Affinity Time-Dependent Density Functional Theory:
    Formalism and Applications to Core-Excited States"
}
\author{
	Kevin Carter-Fenk,\footnote{\href{mailto:carter-fenk@berkeley.edu}{carter-fenk@berkeley.edu}} 
	Leonardo A. Cunha,\footnote{\href{mailto:leonardo.cunha@berkeley.edu}{leonardo.cunha@berkeley.edu}} 
	Juan E. Arias-Martinez,\footnote{\href{mailto:juanes@berkeley.edu}{juanes@berkeley.edu}} 
	and Martin Head-Gordon\footnote{\href{mailto:mhg@cchem.berkeley.edu}{mhg@cchem.berkeley.edu}}\\
	{\em Kenneth S. Pitzer Center for Theoretical Chemistry,}\\
	{\em Department of Chemistry, University of California, Berkeley, CA 94720, USA}\\
	{\em Chemical Sciences Division, Lawrence Berkeley National Laboratory, Berkeley, CA 94720, USA}
}
\date{\today}
\maketitle
\tableofcontents
\clearpage\pagebreak

\section{Linear-Response Time-Dependent Density Functional Theory and its Ion-Orbital Variant}\label{sec:OOTDDFTSI}
The standard time-dependent density functional theory (TDDFT) orbital Hessians are
used for the ``ion-orbital'' TDDFT approach, albeit from a nonstationary
{\em n}-electron reference state that is constructed from the $n-1$-electron molecular orbitals (MOs)
of the core-ionized system.
The usual TDDFT $\mathbf{A}$ and $\mathbf{B}$ matrices take the form,
\begin{equation}\label{eq:OOTDDFT}
    \begin{split}
        A_{ia,ib} &= E^{(n)}\delta_{ab}+F_{ab}^{(n)}-\varepsilon_i^{(n)}\delta_{ab} + (ia|ib)
        - \Chf(ii|ab) + (1-\Chf)(ia|\fxcN|ib)\\
        B_{ia,ib} &= (ia|ib) - \Chf(ib|ai) + (1-\Chf)(ib|\fxcN|ai)
    \end{split}
\end{equation}
where $\fxcN$ is the exchange-correlation kernel, defined as,
\begin{equation}
    \fxcN = \frac{\partial V_{\text{xc}}[\rho^{(n)}]}{\partial\rho^{(n)}}\;,
\end{equation}
and where all quantities denoted with superscript $(n)$ are computed
using the {\em n}-electron density. In the case of IO\nbd-TDDFT, these {\em n}-electron
quantities are constructed from the {\em n}-electron density built from the
unrelaxed $n-1$-electron MOs of the core-ionized system:
\begin{equation}
    P_{\mu\nu}^{(n)} = \sum\limits_{i}^{N} C_{\mu i}^{(n-1)} (C_{\nu i}^{(n-1)})^\ast
\end{equation}

\section{Derivation of the $n-1$-electron Response Kernel}\label{sec:Partials}
In order to correct for the
particle-hole interaction error encountered in the intermediate {\em n}-electron
state obtained after electron addition from the continuum MO,
we take the response of the applied field on the {\em n}-electron state,
which yields the Casida equations for the restricted case,
\begin{equation}
    \begin{split}
        A^{(n)}_{ia,ib} &= E^{(n)}\delta_{ab}+F_{ab}^{(n)}-F_{ii}^{(n)}\delta_{ab} + 2(ia|ib)
        - \Chf(ii|ab) + (1-\Chf)(ia|\fxcN|ib)\\
        B^{(n)}_{ia,ib} &= 2(ia|ib) - \Chf(ib|ai) + (1-\Chf)(ib|\fxcN|ai)
    \end{split}
\end{equation}
and subtract the response of the core orbital with associated the Fock matrix
elements,
\begin{equation}\label{eq:NSFock_SI}
    F_{pq}^{\text{CO}} = F_{pq}^{(n)} - F_{pq}^{(n-1)} = (ii|pq)-\Chf(ip|iq)+(1-\Chf)(p|V_{\text{xc}}^{(n)}-V_{\text{xc}}^{(n-1)}|q) \;,
\end{equation}
where $F_{pq}^{(i)}$ is the core electron's contribution to the Fock matrix of the
{\em n}-electron system.
This form of the Fock matrix
accounts for all couplings between the core-electron components and the
remainder of the {\em n}-electron density. The associated
density matrix is idempotent and contains one electron in the core orbital,
naturally constraining the excitations {\em via} the idemptency condition
such that they can only emerge from core MO $i$.
The response for the corresponding density matrix takes the form,
\begin{equation}\label{eq:NSResponseDerivs_SI}
    \begin{split}
        A^{\text{CO}}_{ia,ib} &= E^{\text{CO}}\delta_{ab} + F_{ab}^{\text{CO}}-F_{ii}^{\text{CO}}\delta_{ab} + \frac{\partial\mathbf{F}_{ia}^{\text{CO}}}{\partial\mathbf{P}_{ib}}\\
        B^{\text{CO}}_{ia,ib} &= \frac{\partial\mathbf{F}_{ai}^{\text{CO}}}{\partial\mathbf{P}_{ib}}
    \end{split}\; ,
\end{equation}
where $E^{\text{CO}} = \tilde{E}(n) - E_0(n-1)$ (the nonstationary {\em n}-electron energy minus the
stationary $n-1$-electron energy of the core ion)
and the partial derivatives yield the final expression for the core-orbital response,
\begin{equation}\label{eq:NSResponse_SI}
    \begin{split}
        A^{\text{CO}}_{ia,ib} &= E^{\text{CO}}\delta_{ab}+F_{ab}^{\text{CO}}-F_{ii}^{\text{CO}}\delta_{ab} + (ia|ib) - \Chf(ii|ab) + (1-\Chf)(ia|\fxcN-\fxcNm|ib)\\
        B^{\text{CO}}_{ia,ib} &= (ia|ib) + (1-\Chf)(ia|\fxcN-\fxcNm|ib)
    \end{split}\; .
\end{equation}
Finally, subtracting the core-orbital part of the response from the full {\em n}-electron response leads to,
\begin{equation}\label{eq:NSCorrectedResponse_SI}
    \begin{split}
        A^{(n)}_{ia,ib} - A^{\text{CO}}_{ia,ib} &= E_0(n-1)\delta_{ab} + F_{ab}^{(n-1)}-F_{ii}^{(n-1)}\delta_{ab} + (ia|ib) + (1-\Chf)(ia|\fxcNm|ib)\\
        B^{(n)}_{ia,ib} - B^{\text{CO}}_{ia,ib} &= (ia|ib) + (1-\Chf)(ia|\fxcNm|ib)
    \end{split}\; .
\end{equation}
We note here that the energy $E^{\text{CO}}$ is equal to the energy of orbital $i$ only for
the exact functional or Hartree-Fock theory, so the explicit form of this energy is never assumed.

\section{Long-Range Self-Interaction Metric}\label{sec:SIEmetric}
Within an ion-orbital {\em ansatz} such as \IOTDA, Eq.~\ref{eq:NSResponse_SI} is suggestive of a metric
that can be used to quantify the degree of long-range self-interaction error ({\em i.e.} the degree of inexact particle-hole interaction)
in approximate density functionals.
Considering only the change in the excitation energy offered by the core-orbital correction,
the total particle-hole interaction error for TDA approximations is,
\begin{equation}\label{eq:metric}
A^{\text{CO}}_{ia,ib} = F_{ab}^{\text{CO}}-F_{ii}^{\text{CO}}\delta_{ab} + (ia|ib) - \Chf(ii|ab) + (1-\Chf)(ia|\fxcN-\fxcNm|ib)
\end{equation}
In Hartree-Fock theory Eq.~\ref{eq:NSFock_SI} implies that $F_{ii}^{\text{CO}}=0$ and that
$F_{ab}^{\text{CO}}+(ia|ib)-(ii|ab)=0$, 
resulting in
a long-range self-interaction error of exactly zero.
It also implies that \IOTDA\ with the HF functional should give equivalent results to
STEX if the nonorthogonality with the {\em n}-electron ground state is not projected out of the STEX Hamiltonian.
This is indeed the case, as \IOTDA\ and EA-TDA produce exactly the same results if the HF functional is used.
If this metric produces a nonzero value, then the density functional approximation being used
incurs some degree of inexact particle-hole interaction and the larger the value of the metric,
the larger the long-range self-interaction error of the functional.

\section{Overlap-Free Transition Dipole Moments}\label{sec:TDMs}
The \EATDA\ spectrum is comprised of states, \{$\Psi_i^a$\}, that are not
orthogonal to the ground state reference, $\Phi_0$, which must be considered
when computing transition properties.
Despite our double-linear-response formalism, we are only interested
the usual transition dipole moments that are observed in one-dimensional x-ray
spectroscopy.
Nonorthogonality between excited state determinants and the ground state
can have severely detrimental effects on transition moments,\cite{WorFeiMan21} but
a simple fix is to subtract the overlap-weighted ground-state dipole moment
from the transition dipole,
\begin{equation}
    \vec{\mu} = \sum\limits_{a} X_i^a \big(\langle\Phi_0|\hat{\mu}|\Psi_i^a\rangle
    - \langle\Phi_0|\hat{\mu}|\Phi_0\rangle \langle\Phi_0|\Psi_i^a\rangle\big) \;,
\end{equation}
where $X_i^a$ are eigenvalues of the Tamm-Dancoff approximated Hermitian eigenvalue equation,
\begin{equation}
    \mathbf{A}\mathbf{X} = \omega\mathbf{X}\; .
\end{equation}
This is equivalent to translating the center of charge of
the molecule to the origin prior to calculating the transition moments.

\section{Additional Data}

\begin{figure}[h!!]
    \centering
    \fig{1.0}{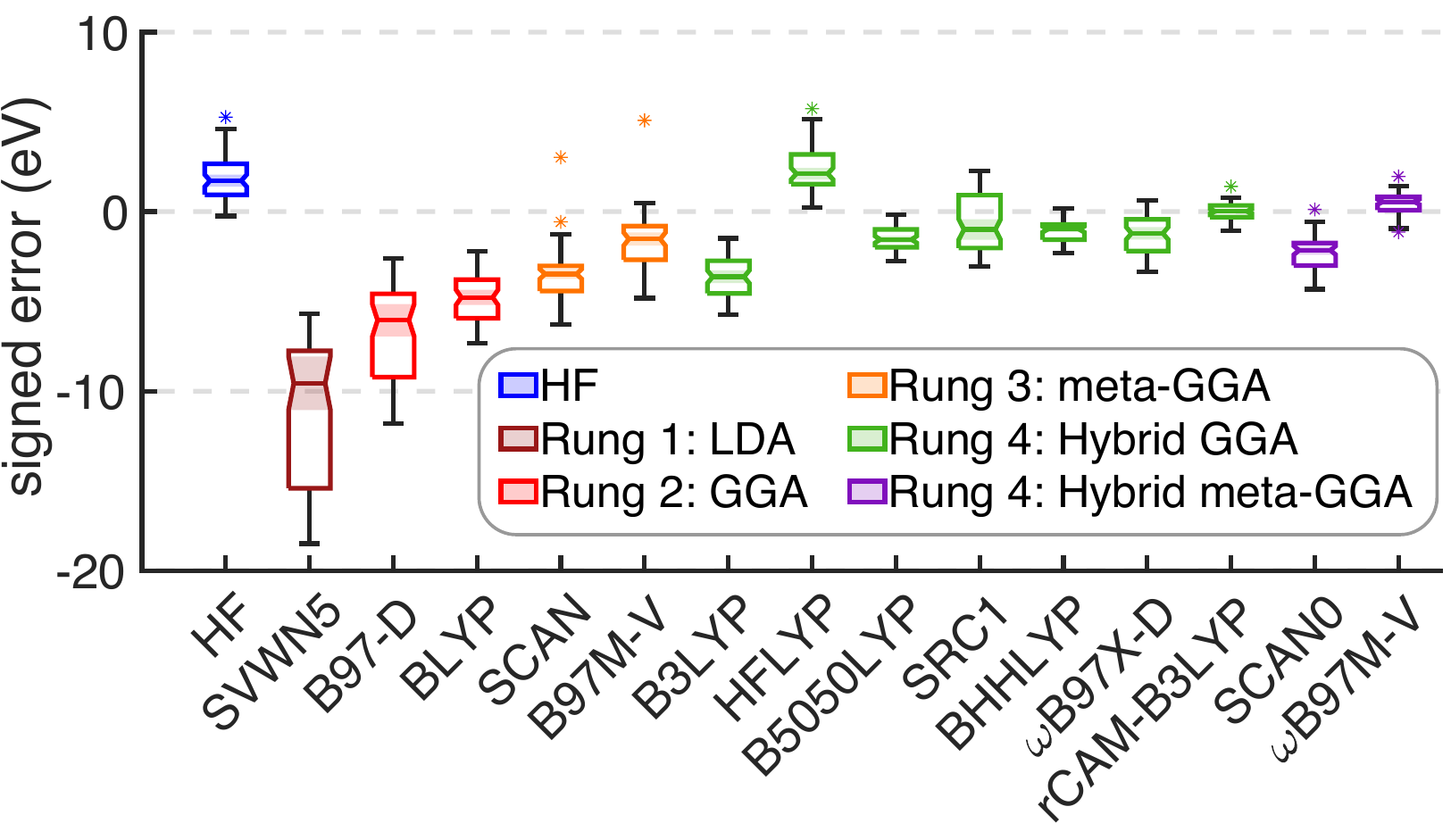}
    \caption{
    \EATDA\
    signed error statistics
    for 65 experimental K-edge transitions (lowest
    energy transition only).
    The aug-pcseg-1 basis was used for H and Br,
    aug-pcX-2 for all other atoms.
    A negative sign indicates an underestimation in the
    excitation energy.
    Upper and lower delimiters indicate maximum
    and minimum errors, respectively. Upper and
    lower bounds of each box are the upper and lower
    quartiles, respectively. Median absolute errors
    are indicated by horizontal lines and
    overlapping notches
    identify statistical similarities between
    distributions to the 95\% confidence level.
    Outliers are indicated by asterisks.
    }
    \label{fig:MSE}
\end{figure}

\begin{figure}
    \centering
    \fig{1.0}{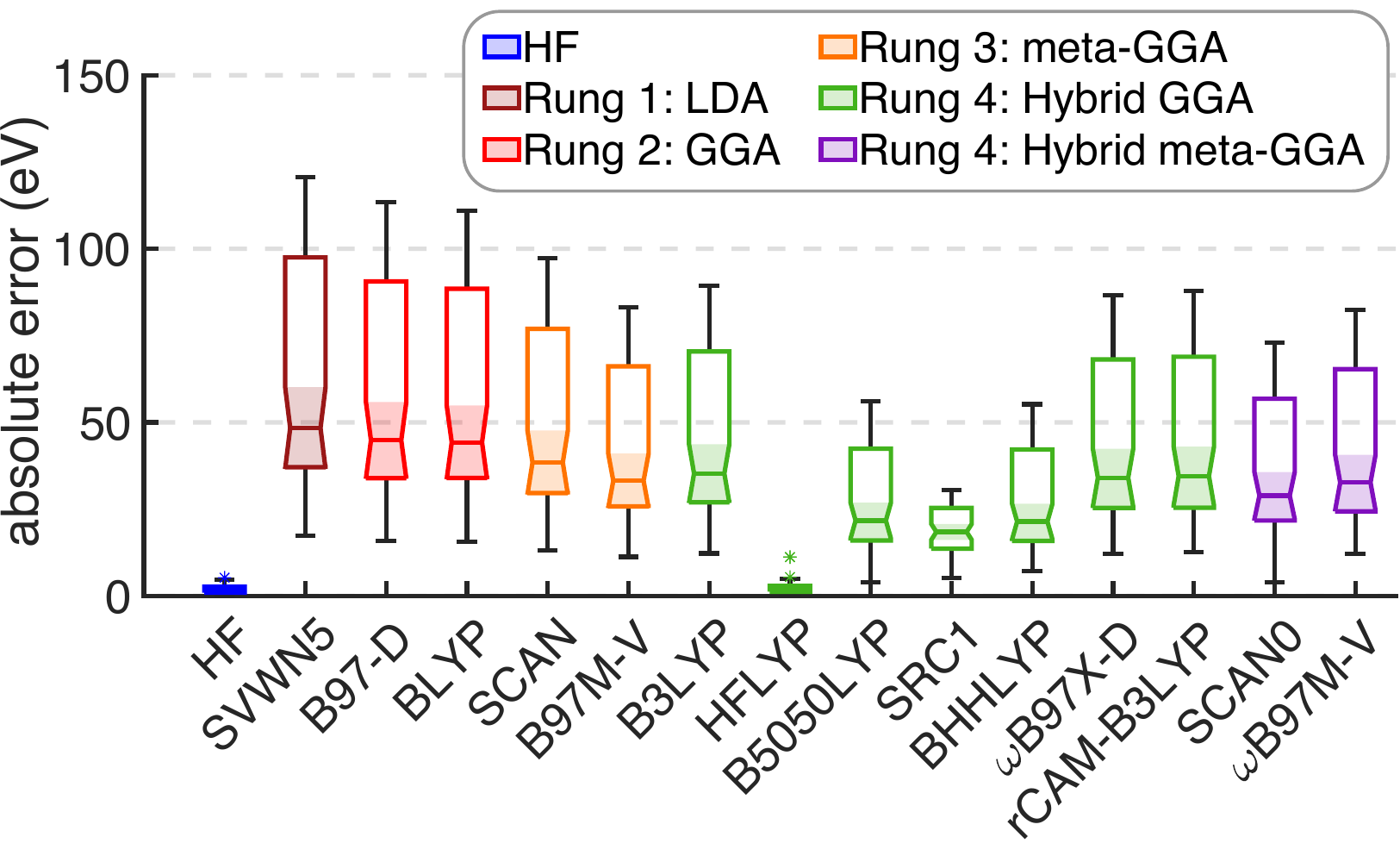}
    \caption{
    \IOTDA\
    absolute error statistics
    for 65 experimental K-edge transitions (lowest
    energy transition only).
    The aug-pcseg-1 basis was used for H and Br,
    aug-pcX-2 for all other atoms.
    Upper and lower delimiters indicate maximum
    and minimum errors, respectively. Upper and
    lower bounds of each box are the upper and lower
    quartiles, respectively. Median absolute errors
    are indicated by horizontal lines and
    overlapping notches
    identify statistical similarities between
    distributions to the 95\% confidence level.
    Outliers are indicated by asterisks.
    }
    \label{fig:OOTDAerrors}
\end{figure}

\begin{figure}
    \centering
    \fig{1.0}{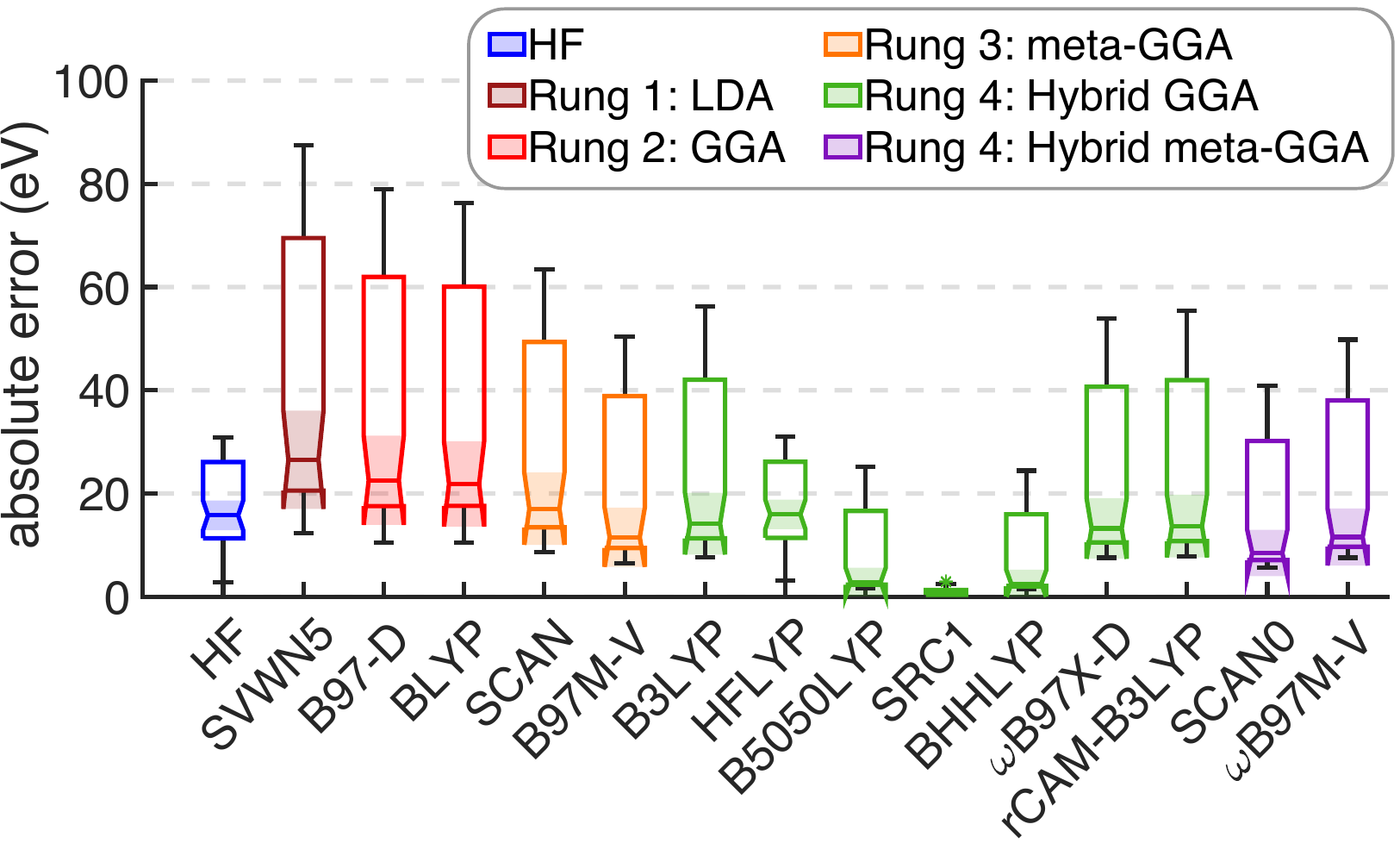}
    \caption{
    Standard TDA
    absolute error statistics
    for 65 experimental K-edge transitions (lowest
    energy transition only).
    The aug-pcseg-1 basis was used for H and Br,
    aug-pcX-2 for all other atoms.
    Upper and lower delimiters indicate maximum
    and minimum errors, respectively. Upper and
    lower bounds of each box are the upper and lower
    quartiles, respectively. Median absolute errors
    are indicated by horizontal lines and
    overlapping notches
    identify statistical similarities between
    distributions to the 95\% confidence level.
    Outliers are indicated by asterisks.
    }
    \label{fig:LRTDAerrors}
\end{figure}

\begin{figure}
    \centering
    \fig{1.0}{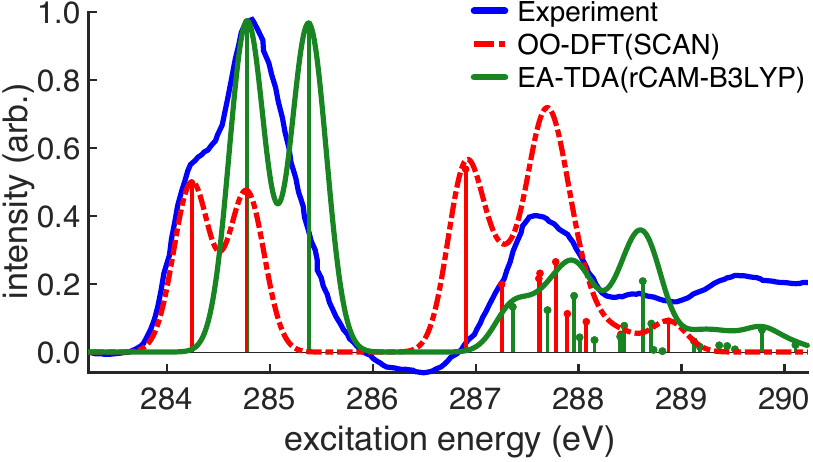}
    \caption{
    {\em trans}-butadiene K-edge
    X-ray absorption spectra for
    \EATDA\ and \OODFT\
    juxtaposed against experimental data
    from Ref.~\citen{SodBri85}.
    The third major peak (intensity of 0.6)
    in the \OODFT\ spectrum corresponds to an
    optically dark 1s$\rightarrow$3s transition
    and is absent in the \EATDA\ spectrum.
    The aug-pcX-2
    and aug-pcseg-1 basis
    sets were used for C and H, respectively.
    }
    \label{fig:butadiene}
\end{figure}

\begin{table}
\caption{Difference between \EATDA(HF) and STEX on 132 K-edge transitions: Be--N}\label{table:EATDAvSTEX_1}
\begin{center}
\scalebox{0.8}{
\begin{tabular}{lcl ......}
\hline\hline
\multirow{2}{*}{Species} & \multirow{2}{*}{Atom} & \multirow{2}{*}{Transition} & \mc{2}{c}{STEX$^a$} & \mc{2}{c}{\EATDA$^a$}
& \multirow{2}{*}{$\Delta$Energy} & \multirow{2}{*}{$\Delta$Strength}\\ \cline{4-5} \cline{6-7}
& & & \mc{1}{c}{Energy} & \mc{1}{c}{Strength} & \mc{1}{c}{Energy} & \mc{1}{c}{Strength} & \\
\hline
Be$^b$  & 	Be  &  1s$\ra$2p  &  115.814  &  9.08$\rm E$-02  &  115.814  & 9.08$\rm E$-02 &  0.000  &  0.00\\
CH$_4$$^c$  & 	C  &  1s$\ra$3s  &  287.303  &  0.00  &  287.323  & 0.00 &  0.019  &  0.00\\
CH$_4$$^c$  & 	C  &  1s$\ra$3p  &  288.441  &  6.30$\rm E$-03  &  288.513  & 6.30$\rm E$-03 &  0.072  &  1.00$\rm E$-08\\
C$_2$H$_2$$^d$  & 	C  &  1s$\ra\pi^\ast$  &  287.219  &  3.81$\rm E$-02  &  287.225  & 3.81$\rm E$-02 &  0.006  &  0.00\\
C$_2$H$_2$$^d$  & 	C  &  1s$\ra$3s  &  288.444  &  4.55$\rm E$-04  &  288.461  & 4.79$\rm E$-04 &  0.016  &  2.45$\rm E$-05\\
C$_2$H$_2$$^d$  & 	C  &  1s$\ra$3p  &  289.492  &  9.82$\rm E$-04  &  289.531  & 9.82$\rm E$-04 &  0.039  &  0.00\\
C$_2$H$_4$$^d$  & 	C  &  1s$\ra\pi^\ast$  &  286.419  &  4.27$\rm E$-02  &  286.428  & 4.28$\rm E$-02 &  0.010  &  1.03$\rm E$-04\\
C$_2$H$_4$$^d$  & 	C  &  1s$\ra$3s  &  287.669  &  1.89$\rm E$-03  &  287.695  & 1.99$\rm E$-03 &  0.026  &  9.65$\rm E$-05\\
C$_2$H$_4$$^d$  & 	C  &  1s$\ra$3p  &  288.263  &  3.48$\rm E$-03  &  288.304  & 3.65$\rm E$-03 &  0.041  &  1.70$\rm E$-04\\
C$_2$H$_6$$^d$  & 	C  &  1s$\ra$3s  &  287.465  &  2.45$\rm E$-03  &  287.487  & 2.52$\rm E$-03 &  0.022  &  7.09$\rm E$-05\\
C$_2$H$_6$$^d$  & 	C  &  1s$\ra$3p  &  288.334  &  4.73$\rm E$-03  &  288.388  & 5.19$\rm E$-03 &  0.054  &  4.60$\rm E$-04\\
C$_6$H$_6$$^d$  & 	C  &  1s$\ra\pi^\ast$  &  286.837  &  4.02$\rm E$-02  &  286.835  & 4.02$\rm E$-02 &  -0.003  &  -4.17$\rm E$-05\\
C$_6$H$_6$$^d$  & 	C  &  1s$\ra$3s  &  287.812  &  2.16$\rm E$-03  &  287.784  & 2.05$\rm E$-03 &  -0.028  &  -1.06$\rm E$-04\\
C$_6$H$_6$$^d$  & 	C  &  1s$\ra$3p  &  288.354  &  9.21$\rm E$-04  &  288.320  & 9.31$\rm E$-04 &  -0.034  &  1.06$\rm E$-05\\
H$_2$CO$^e$  & 	C  &  1s$\ra\pi^\ast$  &  288.041  &  5.94$\rm E$-02  &  288.048  & 5.95$\rm E$-02 &  0.007  &  6.33$\rm E$-05\\
H$_2$CO$^e$  & 	C  &  1s$\ra$3s  &  291.305  &  4.18$\rm E$-03  &  291.309  & 4.27$\rm E$-03 &  0.004  &  8.80$\rm E$-05\\
H$_2$CO$^e$  & 	C  &  1s$\ra$3p (b$_2$)  &  292.189  &  9.88$\rm E$-03  &  292.219  & 1.01$\rm E$-02 &  0.030  &  2.35$\rm E$-04\\
H$_2$CO$^e$  & 	C  &  1s$\ra$3p (b$_1$)  &  292.429  &  2.56$\rm E$-05  &  292.450  & 2.85$\rm E$-05 &  0.022  &  2.87$\rm E$-06\\
HFCO$^f$  & 	C  &  1s$\ra\pi^\ast$  &  290.804  &  7.08$\rm E$-02  &  290.808  & 7.09$\rm E$-02 &  0.004  &  6.52$\rm E$-05\\
HFCO$^f$  & 	C  &  1s$\ra$3s  &  294.246  &  8.72$\rm E$-03  &  294.250  & 8.74$\rm E$-03 &  0.004  &  2.74$\rm E$-05\\
HFCO$^f$  & 	C  &  1s$\ra$3p  &  295.192  &  1.49$\rm E$-03  &  295.197  & 1.52$\rm E$-03 &  0.005  &  3.06$\rm E$-05\\
HCOOH$^g$  & 	C  &  1s$\ra\pi^\ast$  &  290.529  &  7.06$\rm E$-02  &  290.533  & 7.07$\rm E$-02 &  0.004  &  8.80$\rm E$-05\\
HCOOH$^g$  & 	C  &  1s$\ra$3s  &  293.292  &  5.71$\rm E$-03  &  293.302  & 5.91$\rm E$-03 &  0.010  &  2.00$\rm E$-04\\
HCOOH$^g$  & 	C  &  1s$\ra$3p  &  293.592  &  2.76$\rm E$-03  &  293.602  & 2.68$\rm E$-03 &  0.009  &  -7.42$\rm E$-05\\
HCN$^h$  & 	C  &  1s$\ra\pi^\ast$  &  288.098  &  4.63$\rm E$-02  &  288.103  & 4.64$\rm E$-02 &  0.005  &  3.99$\rm E$-05\\
C$_2$N$_2$$^h$  & 	C  &  1s$\ra\pi_{\rm u}^\ast$  &  288.103  &  3.44$\rm E$-02  &  288.103  & 3.44$\rm E$-02 &  0.000  &  -1.00$\rm E$-08\\
C$_2$N$_2$$^h$  & 	C  &  1s$\ra$3s  &  292.167  &  2.71$\rm E$-04  &  292.166  & 2.82$\rm E$-04 &  -0.001  &  1.13$\rm E$-05\\
C$_2$N$_2$$^h$  & 	C  &  1s$\ra\pi_{\rm g}^\ast$\slash 3p  &  293.187  &  4.72$\rm E$-03  &  293.187  & 4.72$\rm E$-03 &  0.000  &  0.00\\
CO$^i$  & 	C  &  1s$\ra\pi^\ast$  &  289.125  &  7.77$\rm E$-02  &  289.125  & 7.77$\rm E$-02 &  0.000  &  0.00\\
CO$^j$  & 	C  &  1s$\ra$3s\slash$\sigma$  &  294.131  &  3.67$\rm E$-03  &  294.127  & 3.68$\rm E$-03 &  -0.004  &  1.04$\rm E$-05\\
CO$^j$  & 	C  &  1s$\ra$3p\slash$\pi$  &  295.025  &  4.08$\rm E$-03  &  295.025  & 4.08$\rm E$-03 &  0.000  &  0.00\\
CO$_2$$^k$  & 	C  &  1s$\ra\pi_{\rm u}^\ast$  &  292.941  &  8.25$\rm E$-02  &  292.941  & 8.25$\rm E$-02 &  0.000  &  1.00$\rm E$-08\\
CO$_2$$^l$  & 	C  &  1s$\ra$3s  &  295.276  &  0.00  &  295.269  & 0.00 &  -0.007  &  0.00\\
CO$_2$$^l$  & 	C  &  1s$\ra$3p  &  297.228  &  1.36$\rm E$-03  &  297.228  & 1.36$\rm E$-03 &  0.000  &  0.00\\
MeOH$^m$  & 	C  &  1s$\ra$3s  &  289.026  &  3.77$\rm E$-03  &  289.044  & 3.77$\rm E$-03 &  0.018  &  3.69$\rm E$-06\\
butadiene$^n$  & 	C(t)  &  1s$\ra\pi^\ast$  &  286.051  &  3.68$\rm E$-02  &  286.056  & 3.68$\rm E$-02 &  0.005  &  3.18$\rm E$-05\\
butadiene$^n$  & 	C(c)  &  1s$\ra\pi^\ast$  &  286.715  &  3.68$\rm E$-02  &  286.717  & 3.68$\rm E$-02 &  0.003  &  -7.55$\rm E$-06\\
furan$^o$  & 	C (3 or 4)  &  1s$\ra\pi^\ast$  &  287.490  &  3.16$\rm E$-02  &  287.495  & 3.16$\rm E$-02 &  0.004  &  3.98$\rm E$-05\\
furan$^o$  & 	C (2 or 5)  &  1s$\ra\pi^\ast$  &  288.160  &  4.24$\rm E$-02  &  288.164  & 4.24$\rm E$-02 &  0.003  &  2.97$\rm E$-05\\
glycine$^p$  & 	C(CO)  &  1s$\ra\pi^\ast$  &  290.885  &  7.32$\rm E$-02  &  290.885  & 7.32$\rm E$-02 &  0.000  &  2.95$\rm E$-05\\
glycine$^p$  & 	C(sp3)  &  1s$\ra\sigma^\ast$  &  289.222  &  4.10$\rm E$-03  &  289.233  & 4.18$\rm E$-03 &  0.011  &  8.60$\rm E$-05\\
HCN$^h$  & 	N  &  1s$\ra\pi^\ast$  &  400.857  &  4.29$\rm E$-02  &  400.862  & 4.30$\rm E$-02 &  0.005  &  5.14$\rm E$-05\\
NH$_3$$^c$  & 	N  &  1s$\ra$3s  &  401.222  &  3.36$\rm E$-03  &  401.226  & 3.37$\rm E$-03 &  0.003  &  1.36$\rm E$-05\\
NH$_3$$^c$  & 	N  &  1s$\ra$3p  &  402.702  &  8.56$\rm E$-03  &  402.737  & 8.56$\rm E$-03 &  0.035  &  0.00\\
NH$_3$$^c$  & 	N  &  1s$\ra$3p  &  403.387  &  5.99$\rm E$-03  &  403.523  & 5.99$\rm E$-03 &  0.136  &  -4.74$\rm E$-06\\
N$_2$$^i$  & 	N  &  1s$\ra\pi^\ast$  &  402.252  &  5.53$\rm E$-02  &  402.252  & 5.53$\rm E$-02 &  0.000  &  0.00\\
N$_2$O$^l$  & 	N(t)  &  1s$\ra\pi^\ast$  &  402.300  &  4.61$\rm E$-02  &  402.300  & 4.61$\rm E$-02 &  0.000  &  0.00\\
N$_2$O$^l$  & 	N(t)  &  1s$\ra$3s\slash$\sigma$  &  405.630  &  1.68$\rm E$-03  &  405.616  & 1.69$\rm E$-03 &  -0.014  &  1.10$\rm E$-05\\
N$_2$O$^l$  & 	N(t)  &  1s$\ra$3p\slash$\pi$  &  407.377  &  1.85$\rm E$-03  &  407.377  & 1.85$\rm E$-03 &  0.000  &  0.00\\
N$_2$O$^l$  & 	N(c)  &  1s$\ra\pi^\ast$  &  406.062  &  5.99$\rm E$-02  &  406.062  & 5.99$\rm E$-02 &  0.000  &  -1.00$\rm E$-08\\
N$_2$O$^l$  & 	N(c)  &  1s$\ra$3s\slash$\sigma$  &  410.517  &  2.11$\rm E$-04  &  410.516  & 2.15$\rm E$-04 &  -0.002  &  3.71$\rm E$-06\\
N$_2$O$^l$  & 	N(c)  &  1s$\ra$3p\slash$\sigma$  &  412.007  &  1.34$\rm E$-04  &  412.007  & 1.34$\rm E$-04 &  0.000  &  0.00\\
C$_2$N$_2$$^h$  & 	N  &  1s$\ra\pi_{\rm u}$  &  400.150  &  3.62$\rm E$-02  &  400.150  & 3.62$\rm E$-02 &  0.000  &  0.00\\
C$_2$N$_2$$^h$  & 	N  &  1s$\ra$3s  &  404.379  &  5.90$\rm E$-05  &  404.376  & 5.92$\rm E$-05 &  -0.004  &  1.80$\rm E$-07\\
C$_2$N$_2$$^h$  & 	N  &  1s$\ra\pi_{\rm g}$\slash 3p  &  405.526  &  3.47$\rm E$-04  &  405.526  & 3.47$\rm E$-04 &  0.000  &  0.00\\
Imidazole$^q$  & 	N (CH=N-CH)  &  1s$\ra\pi^\ast$  &  401.220  &  3.40$\rm E$-02  &  401.222  & 3.40$\rm E$-02 &  0.002  &  3.26$\rm E$-05\\
Imidazole$^q$  & 	N (CH-NH-CH)  &  1s$\ra\pi^\ast$  &  403.650  &  2.43$\rm E$-02  &  403.654  & 2.43$\rm E$-02 &  0.004  &  5.08$\rm E$-05\\
pyrrole$^r$  & 	N  &  1s$\ra\pi^\ast$  &  403.397  &  2.38$\rm E$-02  &  403.402  & 2.39$\rm E$-02 &  0.005  &  1.20$\rm E$-04\\
glycine$^p$  & 	N (NH)  &  1s$\ra\sigma^\ast$  &  401.922  &  2.73$\rm E$-03  &  401.927  & 2.77$\rm E$-03 &  0.004  &  4.35$\rm E$-05\\
glycine$^p$  & 	N (NC)  &  1s$\ra\pi^\ast$  &  402.761  &  4.67$\rm E$-03  &  402.788  & 4.89$\rm E$-03 &  0.028  &  2.24$\rm E$-04\\
\hline\hline
\mc{9}{l}{
	$^a$aug-pcX-2 for non-H and non-Br atoms, aug-pcseg-1 otherwise.
	Data from:
	$^b$Ref.~\citen{KraMar97},
	$^c$Ref.~\citen{SchTroRan93},
	$^d$Ref.~\citen{HitBri77},
	$^e$Ref.~\citen{RemDomPus92},
	$^f$Ref.~\citen{RobIshMcL88},
	$^g$Ref.~\citen{PriRicSim03},
	$^h$Ref.~\citen{HitBri79},
}\\
\mc{9}{l}{
	$^i$Ref.~\citen{SodBri84},
	$^j$Ref.~\citen{HitBri80a},
	$^k$Ref.~\citen{TroKinRea79},
	$^l$Ref.~\citen{PriAvaCor99},
	$^m$Ref.~\citen{HemPiaHei99},
	$^n$Ref.~\citen{SodBri85},
	$^o$Ref.~\citen{DufFlaGiu03},
	$^p$Ref.~\citen{PleFeyRic07},
	$^q$Ref.~\citen{ApeHitGla93},
	$^r$Ref.~\citen{PavHalHen95}
}\\
\end{tabular}}
\end{center}
\end{table}

\begin{table}
\caption{Difference between \EATDA(HF) and STEX on 132 K-edge transitions: O--Ne}\label{table:EATDAvSTEX_2}
\begin{center}
\scalebox{0.8}{
\begin{tabular}{lcl ......}
\hline\hline
\multirow{2}{*}{Species} & \multirow{2}{*}{Atom} & \multirow{2}{*}{Transition} & \mc{2}{c}{STEX$^a$} & \mc{2}{c}{\EATDA$^a$}
& \multirow{2}{*}{$\Delta$Energy} & \multirow{2}{*}{$\Delta$Strength}\\ \cline{4-5} \cline{6-7}
& & & \mc{1}{c}{Energy} & \mc{1}{c}{Strength} & \mc{1}{c}{Energy} & \mc{1}{c}{Strength} & \\
\hline
CO$^b$  & 	O  &  1s$\ra\pi^\ast$  &  534.584  &  3.11$\rm E$-02  &  534.584  & 3.11$\rm E$-02 &  0.000  &  0.00\\
CO$^c$  & 	O  &  1s$\ra$3s\slash$\sigma^\ast$  &  538.608  &  8.71$\rm E$-04  &  538.604  & 8.69$\rm E$-04 &  -0.005  &  -2.35$\rm E$-06\\
CO$^c$  & 	O  &  1s$\ra$3p\slash$\pi^\ast$  &  539.574  &  2.26$\rm E$-05  &  539.570  & 8.40$\rm E$-07 &  -0.004  &  -2.18$\rm E$-05\\
CO$_2$$^d$  & 	O  &  1s$\ra\pi^\ast$  &  536.345  &  2.57$\rm E$-02  &  536.345  & 2.57$\rm E$-02 &  0.000  &  -1.00$\rm E$-08\\
CO$_2$$^d$  & 	O  &  1s$\ra$3s  &  536.619  &  2.57$\rm E$-03  &  536.606  & 2.58$\rm E$-03 &  -0.013  &  1.07$\rm E$-05\\
CO$_2$$^e$  & 	O  &  1s$\ra$3p\slash$\pi_{\rm u}^\ast$  &  538.751  &  3.44$\rm E$-05  &  538.751  & 3.44$\rm E$-05 &  0.000  &  0.00\\
CO$_2$$^e$  & 	O  &  1s$\ra$3p\slash$\sigma^\ast$  &  539.051  &  1.53$\rm E$-03  &  539.050  & 1.52$\rm E$-03 &  -0.001  &  -6.25$\rm E$-06\\
MeOH$^f$  & 	O  &  1s$\ra\sigma^\ast$  &  534.543  &  6.25$\rm E$-03  &  534.547  & 6.27$\rm E$-03 &  0.004  &  1.62$\rm E$-05\\
H$_2$CO$^g$  & 	O  &  1s$\ra\pi^\ast$  &  531.745  &  3.69$\rm E$-02  &  531.747  & 3.69$\rm E$-02 &  0.002  &  4.57$\rm E$-05\\
H$_2$CO$^g$  & 	O  &  1s$\ra$3s  &  535.045  &  5.46$\rm E$-04  &  535.062  & 5.48$\rm E$-04 &  0.017  &  2.13$\rm E$-06\\
H$_2$CO$^g$  & 	O  &  1s$\ra$3p  &  535.978  &  1.16$\rm E$-05  &  535.992  & 4.55$\rm E$-05 &  0.014  &  3.40$\rm E$-05\\
HCFO$^h$  & 	O  &  1s$\ra\pi^\ast$  &  533.115  &  3.49$\rm E$-02  &  533.118  & 3.49$\rm E$-02 &  0.002  &  3.51$\rm E$-05\\
HCFO$^h$  & 	O  &  1s$\ra$3s  &  536.817  &  6.47$\rm E$-04  &  536.822  & 6.22$\rm E$-04 &  0.005  &  -2.47$\rm E$-05\\
HCFO$^h$  & 	O  &  1s$\ra$3p  &  537.142  &  1.93$\rm E$-03  &  537.151  & 1.98$\rm E$-03 &  0.009  &  5.62$\rm E$-05\\
HCOOH$^f$  & 	O (CO)  &  1s$\ra\pi^\ast$  &  533.189  &  3.07$\rm E$-02  &  533.192  & 3.08$\rm E$-02 &  0.003  &  6.54$\rm E$-05\\
HCOOH$^f$  & 	O (OH)  &  1s$\ra\pi^\ast$\slash 3s  &  536.361  &  7.86$\rm E$-03  &  536.360  & 7.87$\rm E$-03 &  -0.001  &  4.09$\rm E$-06\\
H$_2$O$^i$  & 	O  &  1s$\ra$3s  &  534.399  &  7.46$\rm E$-03  &  534.398  & 7.37$\rm E$-03 &  -0.001  &  -9.52$\rm E$-05\\
H$_2$O$^i$  & 	O  &  1s$\ra$3p  &  536.086  &  1.35$\rm E$-02  &  536.110  & 1.33$\rm E$-02 &  0.024  &  -1.95$\rm E$-04\\
N$_2$O$^d$  & 	O  &  1s$\ra\pi^\ast$  &  535.211  &  1.88$\rm E$-02  &  535.211  & 1.88$\rm E$-02 &  0.000  &  1.00$\rm E$-08\\
N$_2$O$^d$  & 	O  &  1s$\ra$3s\slash$\sigma^\ast$  &  537.223  &  3.31$\rm E$-03  &  537.209  & 3.35$\rm E$-03 &  -0.014  &  4.24$\rm E$-05\\
N$_2$O$^d$  & 	O  &  1s$\ra$3p\slash$\pi^\ast$  &  538.930  &  1.57$\rm E$-03  &  538.930  & 1.57$\rm E$-03 &  0.000  &  0.00\\
glycine$^j$  & 	O (CO)  &  1s$\ra\pi^\ast$  &  533.775  &  3.02$\rm E$-02  &  533.774  & 3.02$\rm E$-02 &  -0.001  &  3.70$\rm E$-05\\
glycine$^j$  & 	O (OH)  &  1s$\ra\sigma^\ast$  &  536.307  &  7.00$\rm E$-03  &  536.306  & 7.04$\rm E$-03 &  -0.001  &  3.58$\rm E$-05\\
HCFO$^h$  & 	F  &  1s$\ra\pi^\ast$  &  688.940  &  8.69$\rm E$-03  &  688.943  & 8.70$\rm E$-03 &  0.003  &  1.08$\rm E$-05\\
HF$^k$  & 	F  &  1s$\ra\sigma^\ast$  &  687.533  &  1.34$\rm E$-02  &  687.539  & 1.34$\rm E$-02 &  0.006  &  -3.42$\rm E$-05\\
HF$^k$  & 	F  &  1s$\ra$3p\slash$\sigma^\ast$  &  690.955  &  6.39$\rm E$-03  &  691.037  & 6.94$\rm E$-03 &  0.082  &  5.48$\rm E$-04\\
F$_2$$^k$  & 	F  &  1s$\ra\sigma_{\rm u}$  &  684.087  &  5.18$\rm E$-02  &  684.076  & 5.18$\rm E$-02 &  -0.010  &  -5.61$\rm E$-05\\
F$_2$$^k$  & 	F  &  1s$\ra$3s  &  693.333  &  9.40$\rm E$-04  &  693.333  & 9.14$\rm E$-04 &  -0.001  &  -2.64$\rm E$-05\\
F$_2$$^k$  & 	F  &  1s$\ra$3p  &  693.557  &  2.15$\rm E$-03  &  693.557  & 2.15$\rm E$-03 &  0.000  &  0.00\\
Ne$^{\dagger,l}$  & 	Ne  &  1s$\ra$3s  &  864.931  &  0.00  &  864.923  & 0.00 &  -0.008  &  0.00\\
Ne$^{\dagger,l}$  & 	Ne  &  1s$\ra$3p  &  866.679  &  3.15$\rm E$-03  &  866.679  & 2.48$\rm E$-03 &  0.000  &  -6.71$\rm E$-04\\
\hline\hline
\mc{9}{l}{
	$^a$aug-pcX-2 for non-H and non-Br atoms, aug-pcseg-1 otherwise.
}\\
\mc{9}{l}{
	$^\dagger$Doubly-augmented d-aug-pcX-3 basis due to large basis set incompleteness errors 
}\\
\mc{9}{l}{
	Data from: $^b$Ref.~\citen{SodBri84},
	$^c$Ref.~\citen{HitBri80a},
	$^d$Ref.~\citen{PriAvaCor99},
	$^e$Ref.~\citen{OkaYosSen02},
	$^f$Ref.~\citen{PriRicSim03},
	$^g$Ref.~\citen{RemDomPus92},
	$^h$Ref.~\citen{RobIshMcL88},
	$^i$Ref.~\citen{SchTroRan93},
	$^j$Ref.~\citen{PleFeyRic07},
	$^k$Ref.~\citen{HitBri81},
	$^l$Ref.~\citen{HitBri80b}
}\\
\end{tabular}}
\end{center}
\end{table}

\begin{table}
\caption{Difference between \EATDA(HF) and STEX on 132 K-edge transitions: Si--Cl}\label{table:EATDAvSTEX_3}
\begin{center}
\scalebox{0.8}{
\begin{tabular}{lcl ......}
\hline\hline
\multirow{2}{*}{Species} & \multirow{2}{*}{Atom} & \multirow{2}{*}{Transition} & \mc{2}{c}{STEX$^a$} & \mc{2}{c}{\EATDA$^a$}
& \multirow{2}{*}{$\Delta$Energy} & \multirow{2}{*}{$\Delta$Strength}\\ \cline{4-5} \cline{6-7}
& & & \mc{1}{c}{Energy} & \mc{1}{c}{Strength} & \mc{1}{c}{Energy} & \mc{1}{c}{Strength} & \\
\hline
SiH$_4$$^b$  & 	Si  &  1s$\ra$t2  &  1845.109  &  1.59$\rm E$-03  &  1845.132  & 1.56$\rm E$-03 &  0.023  &  -3.40$\rm E$-05\\
SiH$_4$$^b$  & 	Si  &  1s$\ra$4p  &  1845.860  &  1.02$\rm E$-03  &  1845.882  & 1.18$\rm E$-03 &  0.021  &  1.61$\rm E$-04\\
SiF$_4$$^b$  & 	Si  &  1s$\ra$a1  &  1849.793  &  0.00  &  1849.785  & 0.00 &  -0.008  &  0.00\\
SiF$_4$$^b$  & 	Si  &  1s$\ra$t2  &  1851.330  &  8.73$\rm E$-04  &  1851.330  & 8.73$\rm E$-04 &  0.000  &  0.00\\
SiF$_4$$^b$  & 	Si  &  1s$\ra$4p  &  1852.743  &  4.17$\rm E$-03  &  1852.743  & 4.17$\rm E$-03 &  0.000  &  0.00\\
SiCl$_4$$^b$  & 	Si  &  1s$\ra$a1  &  1848.188  &  0.00  &  1848.171  & 0.00 &  -0.018  &  0.00\\
SiCl$_4$$^b$  & 	Si  &  1s$\ra$t2  &  1849.331  &  5.53$\rm E$-03  &  1849.331  & 5.53$\rm E$-03 &  0.000  &  0.00\\
SiBr$_4$$^b$  & 	Si  &  1s$\ra$a1  &  1846.973  &  0.00  &  1846.993  & 0.00 &  0.020  &  0.00\\
SiBr$_4$$^b$  & 	Si  &  1s$\ra$t2  &  1848.593  &  5.57$\rm E$-03  &  1848.593  & 5.57$\rm E$-03 &  0.000  &  0.00\\
PH$_3$$^c$  & 	P  &  1s$\ra\sigma^\ast$  &  2148.203  &  5.84$\rm E$-04  &  2148.208  & 5.89$\rm E$-04 &  0.005  &  4.81$\rm E$-06\\
PF$_3$$^c$  & 	P  &  1s$\ra\sigma^\ast$  &  2152.812  &  6.60$\rm E$-03  &  2152.812  & 6.60$\rm E$-03 &  0.000  &  0.00\\
PF$_5$$^c$  & 	P  &  1s$\ra\sigma^\ast$  &  2159.064  &  4.39$\rm E$-03  &  2159.064  & 4.39$\rm E$-03 &  0.000  &  0.00\\
POF$_3$$^c$  & 	P  &  1s$\ra\sigma^\ast$  &  2157.270  &  4.95$\rm E$-03  &  2157.270  & 4.95$\rm E$-03 &  0.000  &  1.00$\rm E$-08\\
H$_2$S$^d$  & 	S  &  1s$\ra\sigma^\ast$  &  2475.231  &  1.28$\rm E$-03  &  2475.236  & 1.28$\rm E$-03 &  0.005  &  -3.03$\rm E$-06\\
H$_2$S$^d$  & 	S  &  1s$\ra$Ry  &  2477.071  &  8.87$\rm E$-05  &  2477.183  & 6.19$\rm E$-05 &  0.112  &  -2.68$\rm E$-05\\
CS$_2$$^e$  & 	S  &  1s$\ra$2$\pi_{\rm u}$  &  2473.461  &  2.55$\rm E$-03  &  2473.461  & 2.55$\rm E$-03 &  0.000  &  0.00\\
CS$_2$$^e$  & 	S  &  1s$\ra$3$\sigma_{\rm g}$\slash 3$\sigma_{\rm u}$  &  2476.323  &  1.22$\rm E$-05  &  2476.321  & 1.24$\rm E$-05 &  -0.002  &  1.10$\rm E$-07\\
SF$_4$$^f$  & 	S  &  1s$\ra$b$_2^\ast$  &  2481.921  &  8.83$\rm E$-03  &  2481.921  & 8.83$\rm E$-03 &  0.000  &  0.00\\
SF$_4$$^f$  & 	S  &  1s$\ra$a$_1^\ast$  &  2484.905  &  4.82$\rm E$-03  &  2484.886  & 4.82$\rm E$-03 &  -0.018  &  -6.42$\rm E$-06\\
SF$_4$$^f$  & 	S  &  1s$\ra$b$_1^\ast$  &  2485.866  &  1.01$\rm E$-02  &  2485.866  & 1.01$\rm E$-02 &  0.000  &  0.00\\
SF$_6$$^d$  & 	S  &  1s$\ra\sigma^\ast$ (a$_1$)  &  2487.304  &  0.00  &  2487.267  & 0.00 &  -0.037  &  0.00\\
SF$_6$$^d$  & 	S  &  1s$\ra\sigma^\ast$ (t)  &  2490.756  &  1.56$\rm E$-03  &  2490.756  & 1.56$\rm E$-03 &  0.000  &  1.00$\rm E$-08\\
SO$_2$$^d$  & 	S  &  1s$\ra\sigma^\ast$ (b1)  &  2476.306  &  8.39$\rm E$-03  &  2476.286  & 8.37$\rm E$-03 &  -0.020  &  -1.90$\rm E$-05\\
SO$_2$$^d$  & 	S  &  1s$\ra\sigma^\ast$ (a1)  &  2481.978  &  2.58$\rm E$-03  &  2481.977  & 2.58$\rm E$-03 &  -0.002  &  2.80$\rm E$-07\\
SO$_2$$^d$  & 	S  &  1s$\ra\sigma^\ast$ (b2)  &  2482.856  &  2.31$\rm E$-03  &  2482.856  & 2.31$\rm E$-03 &  0.000  &  0.00\\
SCO$^e$  & 	S  &  1s$\ra$3$\pi$  &  2474.910  &  2.38$\rm E$-03  &  2474.945  & 2.38$\rm E$-03 &  0.034  &  8.91$\rm E$-06\\
SCO$^e$  & 	S  &  1s$\ra$5$\sigma$  &  2476.277  &  8.34$\rm E$-04  &  2476.272  & 8.36$\rm E$-04 &  -0.005  &  1.89$\rm E$-06\\
SCO$^e$  & 	S  &  1s$\ra$6$\sigma$  &  2477.339  &  1.17$\rm E$-03  &  2477.339  & 1.17$\rm E$-03 &  -0.001  &  -3.82$\rm E$-06\\
SF$_5$Cl$^g$  & 	S  &  1s$\ra\sigma^\ast$  &  2484.815  &  9.44$\rm E$-04  &  2484.793  & 9.23$\rm E$-04 &  -0.021  &  -2.05$\rm E$-05\\
SF$_5$Cl$^g$  & 	S  &  1s$\ra\sigma^\ast$  &  2488.638  &  6.67$\rm E$-03  &  2488.631  & 6.71$\rm E$-03 &  -0.007  &  4.04$\rm E$-05\\
SF$_5$Cl$^g$  & 	S  &  1s$\ra\sigma^\ast$  &  2489.517  &  1.12$\rm E$-04  &  2489.516  & 1.09$\rm E$-04 &  -0.001  &  -2.71$\rm E$-06\\
HCl$^h$  & 	Cl  &  1s$\ra$3p $\sigma^\ast$  &  2826.188  &  2.95$\rm E$-03  &  2826.184  & 2.94$\rm E$-03 &  -0.004  &  -2.03$\rm E$-06\\
HCl$^h$  & 	Cl  &  1s$\ra$4s $\sigma$  &  2828.593  &  1.06$\rm E$-03  &  2828.593  & 1.06$\rm E$-03 &  0.000  &  9.70$\rm E$-07\\
HCl$^h$  & 	Cl  &  1s$\ra$4p $\pi$\slash 4p $\sigma$  &  2829.288  &  1.62$\rm E$-04  &  2829.287  & 1.62$\rm E$-04 &  -0.001  &  5.60$\rm E$-07\\
Cl$_2$$^h$  & 	Cl  &  1s$\ra$3p\slash$\sigma_{\rm u}^\ast$  &  2823.962  &  5.58$\rm E$-03  &  2823.959  & 5.58$\rm E$-03 &  -0.003  &  -4.35$\rm E$-06\\
Cl$_2$$^h$  & 	Cl  &  1s$\ra$4p\slash 3d  &  2829.889  &  6.51$\rm E$-04  &  2829.888  & 6.52$\rm E$-04 &  -0.002  &  1.14$\rm E$-06\\
CH$_3$Cl$^i$  & 	Cl  &  1s$\ra$a$_1$  &  2826.649  &  2.29$\rm E$-03  &  2826.652  & 2.31$\rm E$-03 &  0.002  &  1.82$\rm E$-05\\
CH$_3$Cl$^i$  & 	Cl  &  1s$\ra$Ry  &  2827.995  &  1.57$\rm E$-04  &  2828.063  & 1.70$\rm E$-04 &  0.068  &  1.22$\rm E$-05\\
SF$_5$Cl$^g$  & 	Cl  &  1s$\ra\sigma^\ast$  &  2825.235  &  4.49$\rm E$-03  &  2825.233  & 4.49$\rm E$-03 &  -0.002  &  -3.67$\rm E$-06\\
SF$_5$Cl$^g$  & 	Cl  &  1s$\ra$4p  &  2829.231  &  4.91$\rm E$-04  &  2829.230  & 4.93$\rm E$-04 &  -0.001  &  1.88$\rm E$-06\\
CCl$_3$F$^i$  & 	Cl  &  1s$\ra$e  &  2826.865  &  3.38$\rm E$-03  &  2826.863  & 3.37$\rm E$-03 &  -0.002  &  -3.19$\rm E$-06\\
\hline\hline
\mc{9}{l}{
	$^a$aug-pcX-2 for non-H and non-Br atoms, aug-pcseg-1 otherwise.
}\\
\mc{9}{l}{
	Data from: $^b$Ref.~\citen{BodMilNen90},
	$^c$Ref.~\citen{CavJur99},
	$^d$Ref.~\citen{ReyGavBis96},
	$^e$Ref.~\citen{PerLaV84},
	$^f$Ref.~\citen{BodHit87},
	$^g$Ref.~\citen{ReyBodMar92},
	$^h$Ref.~\citen{BodMarRey90},
	$^i$Ref.~\citen{LinCowJac91}
}\\
\end{tabular}}
\end{center}
\end{table}

\begin{table}
\caption{Difference between \EATDA(HF) and STEX transition dipole moments$^a$ on 65 K-edge transitions: Be--N}\label{table:EATDAvSTEX_TDM}
\begin{center}
\scalebox{0.8}{
\begin{tabular}{lcl ...}
\hline\hline
\mc{1}{c}{Species} & \mc{1}{c}{Atom} & \mc{1}{c}{Transition} & \mc{1}{c}{$\Delta\mu_x$} & \mc{1}{c}{$\Delta\mu_y$} & \mc{1}{c}{$\Delta\mu_z$}\\
\hline
Be  & 	Be  &  1s$\ra$2p  &  -2.53$\rm E$-06 & 2.55$\rm E$-03	& 9.42$\rm E$-04\\
CH$_4$  & 	C  &  1s$\ra$3s  & 0.000 & 0.000 & 0.000  \\
C$_2$H$_2$  & 	C  &  1s$\ra\pi^\ast$  & -2.45$\rm E$-03	& 4.87$\rm E$-04	& 0.000\\
C$_2$H$_4$  & 	C  &  1s$\ra\pi^\ast$  &  0.000 & 0.000 & 0.000  \\
C$_2$H$_6$  & 	C  &  1s$\ra$3s  &  0.000	& 3.00$\rm E$-08 &	9.29$\rm E$-06\\
C$_6$H$_6$  & 	C  &  1s$\ra\pi^\ast$  & 0.000 & 0.000 & 0.000\\
H$_2$CO  & 	C  &  1s$\ra\pi^\ast$  & 0.000 & 0.000 & 0.000 \\
HFCO  & 	C  &  1s$\ra\pi^\ast$  & 0.000 & 0.000 & 0.000 \\
HCOOH  & 	C  &  1s$\ra\pi^\ast$  &  0.000 & 0.000 & 0.000 \\
HCN  & 	C  &  1s$\ra\pi^\ast$  & 1.58$\rm E$-03	& -8.48$\rm E$-05 &	0.000 \\
C$_2$N$_2$  & 	C  &  1s$\ra\pi_{\rm u}^\ast$  & -1.49$\rm E$-02 &	3.88$\rm E$-03 &	0.000 \\
CO  & 	C  &  1s$\ra\pi^\ast$  &  8.98$\rm E$-04	& -1.68$\rm E$-05 &	0.000\\
CO$_2$  & 	C  &  1s$\ra\pi_{\rm u}^\ast$  & -1.91$\rm E$-03 &	1.48$\rm E$-03 &	0.000\\
MeOH  & 	C  &  1s$\ra$3s  & -3.82$\rm E$-05 &	-1.26$\rm E$-04 &	0.000 \\
butadiene  & 	C(t)  &  1s$\ra\pi^\ast$  & 0.000 & 0.000 & 0.000 \\
butadiene  & 	C(c)  &  1s$\ra\pi^\ast$  & 7.14$\rm E$-06 &	0.000 &	6.34$\rm E$-05 \\
furan  & 	C (3 or 4)  &  1s$\ra\pi^\ast$  & 0.000 & 0.000 & 0.000 \\
furan  & 	C (2 or 5)  &  1s$\ra\pi^\ast$  & 0.000 & 0.000 & 0.000 \\
glycine  & 	C(CO)  &  1s$\ra\pi^\ast$  & 0.000 & 0.000 & 0.000 \\
glycine  & 	C(sp3)  &  1s$\ra\sigma^\ast$  & 1.83$\rm E$-05 &	-6.31$\rm E$-05	& 0.000 \\
HCN  & 	N  &  1s$\ra\pi^\ast$  &  -1.05$\rm E$-03 &	2.29$\rm E$-05 &	0.000\\
NH$_3$  & 	N  &  1s$\ra$3s  & 0.000	& 0.000	& 2.68$\rm E$-05 \\
N$_2$  & 	N  &  1s$\ra\pi^\ast$  & 8.01$\rm E$-03 &	-9.65$\rm E$-04 &	0.000 \\
N$_2$O  & 	N(t)  &  1s$\ra\pi^\ast$  & -2.00$\rm E$-02	& 1.26$\rm E$-02	& 0.000 \\
N$_2$O  & 	N(c)  &  1s$\ra\pi^\ast$  &  6.48$\rm E$-03 &	-2.43$\rm E$-03 &	0.000 \\
C$_2$N$_2$  & 	N  &  1s$\ra\pi_{\rm u}$  &  -1.69$\rm E$-02	& 5.55$\rm E$-03	& 0.000\\
Imidazole  & 	N (CH=N-CH)  &  1s$\ra\pi^\ast$  & 5.96$\rm E$-06	& 6.12$\rm E$-05	& 0.000 \\
Imidazole  & 	N (CH-NH-CH)  &  1s$\ra\pi^\ast$  & 0.000	& 0.000	& -1.00$\rm E$-08 \\
pyrrole  & 	N  &  1s$\ra\pi^\ast$  &  0.000	& 0.000	& 6.49$\rm E$-05\\
glycine  & 	N (NH)  &  1s$\ra\sigma^\ast$  & 2.70$\rm E$-05 &	-3.74$\rm E$-05 &	0.000 \\
CO  & 	O  &  1s$\ra\pi^\ast$  &  2.04$\rm E$-04	& -4.13$\rm E$-06	& 0.000\\
CO$_2$  & 	O  &  1s$\ra\pi^\ast$  & 5.91$\rm E$-04	& -9.88$\rm E$-05 &	0.000 \\
MeOH  & 	O  &  1s$\ra\sigma^\ast$  &  3.77$\rm E$-05	& -5.72$\rm E$-05 &	0.000\\
H$_2$CO  & 	O  &  1s$\ra\pi^\ast$  & 0.000	& 0.000 & 0.000 \\
HCFO  & 	O  &  1s$\ra\pi^\ast$  & 0.000	& 0.000 & 0.000 \\
HCOOH  & 	O (CO)  &  1s$\ra\pi^\ast$  & 3.13$\rm E$-05	& -4.02$\rm E$-06	& 0.000 \\
HCOOH  & 	O (OH)  &  1s$\ra\pi^\ast$\slash 3s  &  0.000	& 0.000 & 0.000   \\
H$_2$O  & 	O  &  1s$\ra$3p  & 0.000	& 0.000 &	3.03$\rm E$-05  \\
N$_2$O  & 	O  &  1s$\ra$3s\slash$\sigma^\ast$  & 9.17$\rm E$-03 &	-2.35$\rm E$-03	& 0.000 \\
glycine  & 	O (CO)  &  1s$\ra\pi^\ast$  &  2.83$\rm E$-05	& 1.55$\rm E$-05	& 0.000\\
glycine  & 	O (OH)  &  1s$\ra\sigma^\ast$  & 0.000	& 0.000 & 0.000 \\
HCFO  & 	F  &  1s$\ra\pi^\ast$  & 0.000	& 0.000 & 0.000 \\
HF  & 	F  &  1s$\ra\sigma^\ast$  & 0.000	& 0.000 &	7.75$\rm E$-06 \\
F$_2$  & 	F  &  1s$\ra$3s  & 0.000	& 0.000  & -2.10$\rm E$-05 \\
Ne$^\dagger$  & 	Ne  &  1s$\ra$3p  & 0.000	& 0.000 & 0.000 \\
SiH$_4$  & 	Si  &  1s$\ra$t2  & 0.000	& 0.000 & 0.000 \\
SiF$_4$  & 	Si  &  1s$\ra$a1  & 0.000	& 0.000 & 0.000 \\
SiCl$_4$  & 	Si  &  1s$\ra$a1  & 0.000	& 0.000 & 0.000 \\
SiBr$_4$  & 	Si  &  1s$\ra$a1  & 0.000	& 0.000 & 0.000 \\
PH$_3$ & P & 1s$\ra\sigma\ast$ &  0.000	& -2.00$\rm E$-08 &	6.80$\rm E$-07\\
PF$_3$  & 	P  &  1s$\ra\sigma^\ast$  & -8.00$\rm E$-08	& 0.000	& 0.000\\
PF$_5$  & 	P  &  1s$\ra\sigma^\ast$  & 0.000 &	-1.00$\rm E$-08	& 0.000 \\
POF$_3$  & 	P  &  1s$\ra\sigma^\ast$  & 0.000 &	0.000 &	1.41$\rm E$-05 \\
H$_2$S  & 	S  &  1s$\ra\sigma^\ast$  & 0.000 &	0.000	& 5.50$\rm E$-07 \\
CS$_2$  & 	S  &  1s$\ra$2$\pi_{\rm u}$  & 1.89$\rm E$-04	& -1.94$\rm E$-05	& 0.000 \\
SF$_4$  & 	S  &  1s$\ra$b$_2^\ast$  &  0.000	& 0.000 & 0.000\\
SF$_6$  & 	S  &  1s$\ra\sigma^\ast$ (a$_1$)  & 0.000	& 0.000 & 0.000\\
SO$_2$  & 	S  &  1s$\ra\sigma^\ast$ (b1)  &  0.000	& 0.000 & 0.000\\
SCO  & 	S  &  1s$\ra$3$\pi$  & -1.79$\rm E$-04 &	3.43$\rm E$-05	& 0.000 \\
SF$_5$Cl  & 	S  &  1s$\ra\sigma^\ast$  & 0.000 & 0.000 & -3.04$\rm E$-05\\
HCl  & 	Cl  &  1s$\ra$3p $\sigma^\ast$  &  0.000	& 0.000 & -1.59$\rm E$-06\\
Cl$_2$ & 	Cl  &  1s$\ra$4p\slash 3d  & 0.000	& 0.000 & -2.47$\rm E$-06 \\
CH$_3$Cl  & 	Cl  &  1s$\ra$a$_1$  & 0.000	& 0.000 & 2.00$\rm E$-06 \\
SF$_5$Cl  & 	Cl  &  1s$\ra\sigma^\ast$  & 0.000	& 0.000 & -2.33$\rm E$-06 \\
CCl$_3$F  & 	Cl  &  1s$\ra$e  & 0.000	& -2.39$\rm E$-06	& -7.00$\rm E$-08 \\
\hline\hline
\mc{6}{l}{
 $^a$aug-pcX-2 for non-H and non-Br atoms, aug-pcseg-1 otherwise.
 $\Delta\mu_p=\mu_p^{\text{EA-TDA}} - \mu_p^{\text{STEX}}$.
}\\
\mc{6}{l}{
  $^\dagger$Doubly-augmented d-aug-pcX-3 basis due to large basis set incompleteness errors 
}\\
\end{tabular}}
\end{center}
\end{table}

\clearpage
\pagebreak

\providecommand{\refin}[1]{\\ \textbf{Referenced in:} #1}